\documentclass[journal,10pt,twocolumn,twoside]{IEEEtran}
 
\usepackage[dvips]{graphicx}
\usepackage[utf8]{inputenc}
\usepackage[T1]{fontenc}
\usepackage{cite}
\usepackage{diagbox}

\usepackage[left=14mm, right=14mm, top=19mm, bottom=43mm]{geometry}  
\usepackage{amsmath,amssymb,amsfonts,amsthm,mathrsfs}
\usepackage{epsfig,epsf,subfigure,graphicx,graphics}
\usepackage{url,enumerate}
\usepackage{color}
\usepackage{CJK,tikz}
\usepackage{multirow}
\usepackage{algorithm,algorithmicx}
\usepackage{algpseudocode}

\usepackage{subfigure}

\newtheoremstyle{mytheorem}
  {3pt}
  {3pt}
  {\itshape}
  {}
  {\itshape}
  {.}
  {.5em}
  {\thmname{#1}\thmnumber{{ }#2}%
   \thmnote{ {\the\thm@notefont(#3)}}}
\makeatother
\theoremstyle{mytheorem}
\theoremstyle{mytheorem}
\newtheorem{theorem}{Theorem}
\newtheorem*{definition}{Definition}
\newtheorem{lemma}{Lemma}

\begin{document}

\title{Downlink SCMA Codebook Design with Low Error Rate by Maximizing Minimum Euclidean Distance of Superimposed Codewords}


\author{Chinwei~Huang,~\IEEEmembership{Student Member,~IEEE,}~Borching Su,~\IEEEmembership{Member,~IEEE,}~Tingyi~Lin,~and~Yenming~Huang~\IEEEmembership{Member,~IEEE
}
}

\maketitle

\begin{abstract}

Sparse code multiple access (SCMA), as a codebook-based non-orthogonal multiple access (NOMA) technique, has received research attention in recent years. 
The codebook design problem for SCMA has also been studied to some extent since codebook choices are highly related to the system's error rate performance.
In this paper, we approach the SCMA codebook design problem by formulating an optimization problem to maximize the minimum Euclidean distance (MED) of superimposed codewords under power constraints.
While SCMA codebooks with a larger MED are expected to obtain a better BER performance, no optimal SCMA codebook in terms of MED maximization, to the authors' best knowledge, has been reported in the SCMA literature yet.
In this paper, a new iterative algorithm based on alternating maximization with exact penalty is proposed for the MED maximization problem. 
The proposed algorithm, when supplied with appropriate initial points and parameters, achieves a set of codebooks of all users whose MED is larger than any previously reported results. 
A Lagrange dual problem is derived which provides an upper bound of MED of any set of codebooks.
Even though there is still a nonzero gap between the achieved MED and the upper bound given by the dual problem, simulation results demonstrate clear advantages in error rate performances of the proposed set of codebooks over all existing ones not only in AWGN channels but also in some downlink scenarios that fit in 5G/NR applications, making it a good codebook candidate thereof.
The proposed set of SCMA codebooks, however, are not shown to outperform existing ones in uplink channels or in the case where non-consecutive OFDMA subcarriers are used.
The correctness and accuracy of error curves in the simulation results are further confirmed by the coincidences with the theoretical upper bounds of error rates derived for any given set of codebooks.

\end{abstract}


\section{Introduction}
\label{Sec:Introduction}

In the fifth generation (5G) wireless communications and beyond, to enable the massive connectivity and high spectral efficiency for the Internet of Things (IoT) and the Factories of the Future (FoF), non-orthogonal multiple access (NOMA) \cite{NOMA_performance} is considered an important multiple access scheme due to its extended spatial efficiency as opposed to the conventional orthogonal multiple access (OMA), such as orthogonal frequency-division multiplexing (OFDM). 
Among the many existing schemes in NOMA \cite{Dai2015}, SCMA is regarded as a promising multiple access scheme \cite{SCMA_1, SCMA_2, SCMA_3}.
Sparse code multiple access (SCMA) \cite{SCMA_ori} is one kind of code-domain NOMA, which distinguishes multiple users with the aid of codewords \cite{Wang2018}. 
Incoming bits are directly mapped to multi-dimensional codewords of some set of SCMA codebooks, so the codebook design dominates the performance of the SCMA-based NOMA system \cite{SCMA_suboptimal}.

Codebook design for SCMA has been studied extensively in the past few years \cite{Huawei2015, starQAM2015, Peng2017, Alam2017, Zhou2017, Sharma2018, topdown2016, Zhang2016, Chen2020, Deka2020, Mheich2019, Jiang2020, Hasan2020, Cai2016, Lai2020, Vameghestahbanati2019, Yu2018, Wang2019, Lou2018}.
The overall design goal is to find a set of codebooks that result in a good performance, in terms of low error rate or large spectral efficiency, in the scenarios of AWGN, uplink, and downlink fading channels.
One of the major approaches for this goal is to find codebooks that have a large minimum Euclidean distance (MED) \cite{topdown2016, starQAM2015, Peng2017, Alam2017, Wang2019, Chen2020}.
The basic rationale behind the approaches of maximizing MED is that a codebook with a larger MED usually results in better error rate performance and we choose the approach of MED maximization in this paper.
In this regard, pioneering works including Yu et al\cite{Yu2018}, and many others \cite{Cai2016, Zhang2016, Zhou2017, Chen2020} considered multi-stage design approaches by first constructing a mother constellation (MC) and then letting every user apply the mother constellation with different rotation and permutation operations and occupy different resources.
Under this multi-stage design approach, the maximization of MED of mother constellation has been considered as an important issue and has been studied to some extents \cite{topdown2016, starQAM2015, Alam2017}.
While this approach has rather a simple complexity in the optimization problem, the fact that codebooks of all users are tied to a fixed mother constellation implicitly impose extra and probably unnecessary constraints to the choice of codebooks, and may lead to a suboptimal codebook design solution.

More recently, the idea of MED maximization is studied with a newer definition of MED, namely, the MED of superimposed codewords \cite{Peng2017,Wang2019,Chen2020}.
Many previously reported codebook design methods \cite{Zhou2017, Peng2017, Yu2018, Lou2018, Wang2019, Chen2020} have used MED of superimposed codewords as one design KPI.
However, few of the previous works have directly maximized the MED of superimposed codewords with only power constraints and the reason may be the overwhelming complexity while dealing with this non-convex optimization problem. 
Although an MED maximization problem has been formulated in \cite{Peng2017}, the algorithm proposed therein does not guarantee to obtain the optimal point.
In fact, obtaining the set of codebooks with the maximal MED is still an open question today, nor has an upper bound of the maximal MED been known yet.
For convenience, hereafter we use the term ``MED'' as the MED of superimposed codewords, rather than of the mother constellation, throughout this paper.

In this work, we propose to approach the SCMA codebook design problem by maximizing the MED of the designed set of codebooks.
The major contributions of the paper are summarized as follows: 1) A new method is proposed to deal with the non-convex optimization problem based on the exact penalty technique \cite{Demir2015} with an alternating maximization \cite{AM} approach.
2) The aforementioned method achieves a codebook design that has a larger MED than any previously reported design, which also shows the best error rate performances among all existing codebooks in some cases, including the one which fits in the downlink scenario of 5G/NR applications.
3) A theoretical upper bound for MED of any possible codebook designs, that was not known before, is obtained by deriving a Lagrange dual problem of the main problem.
The rest of the paper is organized as follows. 
Section \ref{Sec:SMPM} describes the downlink SCMA system model based on OFDMA. 
In Section \ref{Sec:Max_of_MED}, the  MED  maximization problem is formulated and a corresponding algorithm is proposed.
In Section \ref{Sec:Dual_problem}, the dual problem of the MED maximization problem is derived.
The numerical results are given in Section \ref{Sec:SimResult}.
Finally, conclusions are made in Section \ref{Sec:Conclusion}.


\subsection{Notations}
\label{Sec:Notations}
Boldfaced lower case letters such as $\mathbf{x}$ represent column vectors, boldfaced upper case letters such as $\mathbf{X}$ represent matrices, and italic letters such as $x,X$ represent scalars.
Superscripts as in $\mathbf{X}^T$, $\mathbf{X}^{H}$, and $\mathbf{X}^{-1}$ denote the transpose, transpose-conjugate, and inverse operators, respectively, of a matrix.
The binary set $\{0,1\}$ is denoted by $\mathbb{B}$. 
Given any positive integer $n$,  $\mathcal{Z}_{n}$ stands for the set $\left\{ 1,2,\dots,n \right\}$.
The $n$-dimensional complex, binary, and integer vector spaces are expressed as $\mathbb{C}^{n}$, $\mathbb{B}^{n}$, and $\mathbb{Z}^{n}$, respectively.
The $(n\times \nu)$-dimensional complex, binary, and integer matrix spaces are expressed as $\mathbb{C}^{n\times \nu}$, $\mathbb{B}^{n\times \nu}$, and $\mathbb{Z}^{n\times \nu}$, respectively.
The set of all $n\times n$ Hermitian matrices is denoted by $\mathbb{H}^n$ and the set of all positive semidefinite matrices is denoted by $\mathbb{H}^n_{+}$.
The notation $\mathbf{A}\preceq\mathbf{B}$ means $\mathbf{B}-\mathbf{A}\in\mathbb{H}^n_+$.
Let $\mathbf{1}_n$, $\mathbf{0}_{n}$, $\mathbf{e}_{\nu}^{(n)}$ be the all-one vector, all-zero vector, and $\nu$-th standard unit vector, respectively, of dimension $n$.
Let 
$\mathbf{I}_{n}$ be the 
$n\times n$ identity matrix. 
Operators $\left \| \cdot  \right \|_p$, $\mathrm{tr}(\cdot)$, $\mathrm{vec}(\cdot)$, $\circ$,  $\otimes$, and $\times$ denote $\ell_p$-norm, trace, vectorization, Hadamard product, Kronecker product, and Cartesian, respectively.
For some events $A,B$, the probability of $A$ and the conditional probability of $A$ given $B$ are denoted by $\mathrm{Pr}\{ A \}$ and $\mathrm{Pr}\{ A| B\}$, respectively. 
For some random variables $X$ and $Y$, the conditional probability density function of $X$ given $Y$ is denoted by $f(X|Y)$. 
The expectation of $X$ is denoted by $\mathrm{E}\{X\}$.
Throughout the paper, we adopt one-based indexing.
For some vector $\mathbf{x}$ and matrix $\mathbf{X}$, the $i$-th entry of $\mathbf{x}$ and the $(i,j)$-th entry of $\mathbf{X}$ are denoted by $[\mathbf{x}]_{i}$ and $[\mathbf{X}]_{i,j}$, respectively. 


\section{System Model}
\label{Sec:SMPM}

\subsection{Downlink SCMA System Based on OFDMA}
\label{downlink_SCMA_system_based_on_OFDMA}
\begin{figure*}[t]
\centering \centerline{
\includegraphics[width=1.0\textwidth,clip]{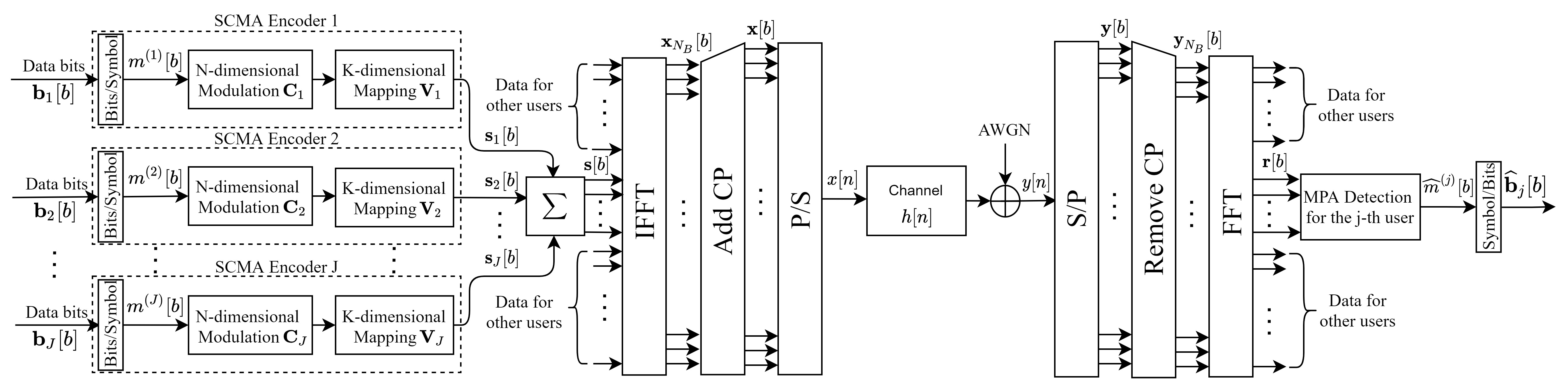}}
\caption{Downlink SCMA system model based on OFDMA.}
\label{SystemModel}
\end{figure*} 

\label{Sec:downlinkSCMAsystem}
We consider downlink SCMA transmission on top of an underlying orthogonal frequency division multiple access (OFDMA) system since SCMA multiplexed symbols need to be transmitted over orthogonal resource elements \cite{HuaweiDownlink2014}.
The block diagram of such a downlink SCMA system is shown in Fig. \ref{SystemModel} where $J$, $K$, and $N_B$ represent the number of users, the number of resources, and the FFT size in the underlying OFDMA systems, respectively. 
The message carrying the data bits of the $b$-th block transmission (i.e., the $b$-th OFDM symbol) for the $j$-th user is encoded and mapped into $K$-dimensional symbols $\mathbf{s}_{j}[b]$ by the $j$-th specific SCMA encoder, and then the sum of these codewords, ${\bf s}[b]$, is transmitted over $K$ orthogonal resources, which are $K$ consecutive subcarriers starting from the $i_{sub}$-th subcarrier of the $b$-th OFDM block. 
Other subcarriers of the OFDM block may contain data from other users.
The result of the IFFT operation, ${\bf x}_{N_B}[b]$, is further added a cyclic prefix of length $N_{CP}$ to obtain ${\bf x}[b]$ before being sent to the channel with a finite impulse response (FIR), characterized by $h[n]$, whose order is upper bounded by $L-1$. 
Following the standard OFDMA receiver, at the output of the FFT operation, excluding subcarriers containing data from other users, the receiver observes the signal of the $b$-th block transmission as
\begin{align}
\label{received_signal}
\mathbf{r}[b] = \mathrm{diag}(\mathbf{h}_{sub}^f)\mathbf{s}[b]+\mathbf{n}[b]
\end{align} 
where $\mathbf{h}_{sub}^f=\left[[\mathbf{h}^f]_{i_{sub}}~ [\mathbf{h}^f]_{i_{sub}+1}~ \cdots ~[\mathbf{h}^f]_{i_{sub}+K-1}\right]^T\in \mathbb{C}^{K}$ is the subvector of the frequency-domain channel gain vector $\mathbf{h}^f\in \mathbb{C}^{N_B}$, $\mathbf{s}[b] = \sum _{j=1}^{J}\mathbf{s}_{j}[b]\in \mathbb{C}^{K}$ is the transmitted superimposed codeword of the $b$-th block transmission, and $\mathbf{n}[b]\sim \mathcal{CN}(\mathbf{0}_K, N_0 \mathbf{I}_K)$ is the additive white Gaussian noise.
The channel gain vector $\mathbf{h}^f$ is the discrete Fourier transform of vector $\left[\mathbf{h}^T \enspace \mathbf{0}^T_{N_B-L} \right]^T\in \mathbb{C}^{N_B}$, where $\mathbf{h}=\left[h[0]~h[1]~\cdots~h[L-1]\right]^T\in \mathbb{C}^{L}$ is the channel vector which represents the channel impulse response as FIR filter of length $L$. 

In 5G/NR applications, multipath channels are extensively considered \cite{3GPP_38802, 3GPP_TS38211} and therefore the orthogonality of resources in SCMA systems needs to be ensured by an OMA technique, usually as OFDMA \cite{HuaweiDownlink2014, 3GPP_TR38812}.
When the popular OMA technique, OFDMA, is considered for the provision of orthogonal resources, consecutive subcarriers are usually chosen to form a resource block rather than separate subcarriers. (The interested readers can refer to Section 4.4.4 in \cite{3GPP_TS38211}.)  Based on these facts, in order to fit the scenario of 5G/NR applications where SCMA is widely discussed \cite{3GPP_TR38812}, we, therefore, consider the SCMA system based on OFDMA with consecutive subcarriers chosen as orthogonal resources as what we set in the last paragraph.


\subsection{SCMA Encoder}\label{Sec:SCMA_Encoder}

An SCMA encoder for the $j$-th user can be regarded as a function defined as
\begin{align}
\label{SCMA_encoder}
f_j:\mathbb{B}^{\mathrm{log}_{2}M} \rightarrow \mathcal{S}_{j}
\end{align} 
where $\mathcal{S}_{j}\subset \mathbb{C}^K$ is the codebook of the $j$-th user with cardinality $\left |  \mathcal{S}_{j} \right |=M$, i.e., ${\cal S}_j$ contains $M$ codewords. 
We require $M$ to be a power of two so that each codeword in $\mathcal{S}_j$ represents $\log_2 M$ bits of information.
For notational convenience, we say that $m^{(j)}\in {\cal Z}_M$ is an \emph{SCMA symbol} according to the vector of data bits ${\bf b}_j \in \mathbb{B}^{\log_2 M}$ from the $j$-th user:
$ m^{(j)} = 1+\sum_{i=1}^{\log_2 M} 2^{i-1} [{\bf b}_j]_i.$
Each SCMA symbol $m^{(j)}$ maps to a $K$-dimensional complex codeword $\mathbf{s}_{j} \in \mathcal{S}_{j}$, which is a sparse vector with $N$ non-zero entries, and $K > N$.
Following \cite{Peng2017, Rueibin2019}, the $j$-th user's codebook is chosen as
\begin{equation}
\label{eq:S_j}    
 {\cal S}_j = \{{\bf V}_j{\bf C}_j{\bf e}_{m}^{(M)} ~|~ m\in {\cal Z}_M \}\end{equation}
where $\mathbf{C}_j\in\mathbb{C}^{N\times M}$ and $\mathbf{V}_{j}\in\mathbb{B}^{K\times N}$ are the constellation matrix and the mapping matrix of the $j$-th user, respectively. 
The mapping matrix $\mathbf{V}_{j}$ is obtained by removing $K-N$ columns from $\mathbf{I}_{K}$. 
Now, the codeword in $\mathcal{S}_j$ selected by the $j$-th user for the $b$-th block transmission can be expressed as
\begin{align}\label{bit2codeword}
\mathbf{s}_{j}[b] = f_j(\mathbf{b}_j[b])=\mathbf{V}_{j} \mathbf{C}_j \mathbf{e}_{m^{(j)}[b]}^{(M)}\in \mathbb{C}^{K}
\end{align}
where $\mathbf{b}_j[b]\in \mathbb{B}^{\mathrm{log}_{2}M}$ is the vector of the given data bits and $m^{(j)}[b]\in\mathcal{Z}_M$ is the corresponding SCMA symbol. 
All codewords chosen by the $J$ users will be summed up together to form a \emph{superimposed codeword} before being assigned to orthogonal resources:
$$ {\bf s}[b] = \sum_{j=1}^J {\bf s}_j[b].$$
A superimposed codeword is  determined by all SCMA symbols from $J$ users, collectively a vector in ${\cal Z}_M^J$ which we refer to as the \emph{multiplexed symbol}.
There are in total $M^J$ multiplexed symbols, and we denote the $k$-th multiplexed symbol, $k\in{\cal Z}_{M^J}$, by ${\bf m}_k$ defined by 
\begin{equation*}
{\bf m}_k = [k_1, k_2, ..., k_J]\in \mathcal{Z}_M^J
\end{equation*}where $k_1,..., k_J\in{\cal Z}_M$ are the unique integers that satisfy
\begin{equation}
k = 1+\sum_{j=1}^J (k_j - 1) M^{j-1}  \in {\cal Z}_{M^J}.\label{eq:multiplexed_symbol_index}
\end{equation}
Note that the set $\mathcal{Z}_{M}^J\subset \mathbb{Z}^J$ stands for the Cartesian product of $J$ identical sets, $\mathcal{Z}_{M}$.
The set of all SCMA superimposed codewords can be expressed as
\begin{align}
    \label{eq:superimposed_codeword}
    {\cal S} = \left\{\sum_{j=1}^{J} {\bf V}_j{\bf C}_j{\bf e}_{k_j}^{(M)} ~\bigg|~  k \in \mathcal{Z}_{M^J}\right\},
\end{align}
where $k\in{\cal Z}_{M^J}$ and $k_j\in{\cal Z}_M$ follow the relationship defined in \eqref{eq:multiplexed_symbol_index}.

Since each of the $J$ users has a distinct mapping matrix ${\bf V}_j$, it is obvious that $J\leq \binom{K}{N}$ must hold for any given $K$ and $N$.
Furthermore, a loading factor is defined as $\lambda = J/K$ \cite{Zhou2017} that directly determines the spectral efficiency of SCMA. 
It must satisfy $\lambda > 1$ for SCMA to be more spectral efficient than conventional OMA, and it is well known that this implies $K\geq 4$, $2 \leq N\leq K-2$, and $J>K$ \cite{Zhou2017}.
In order to achieve the maximum sparsity, $N=2$ is often chosen.
In this paper, we choose to study the simplest case by taking $K=4, N=2$, and $M=4$, which is also the most studied case in the literature.
Since $J\leq \binom{K}{N}$, we take $J = 6$ and set the mapping matrices $\mathbf{V}_1, \mathbf{V}_2, \dots, \mathbf{V}_J$ as follows:
\begin{align}\label{mapping_matrix}
\mathbf{V}_1=\begin{bmatrix}
1 & 0\\
0 & 1\\
0 & 0\\
0 & 0
\end{bmatrix}
\mathbf{V}_2=\begin{bmatrix}
0 & 0\\ 
0 & 0\\ 
1 & 0\\ 
0 & 1
\end{bmatrix}
\mathbf{V}_3=\begin{bmatrix}
1 & 0\\ 
0 & 0\\ 
0 & 1\\ 
0 & 0
\end{bmatrix}~ \nonumber \\
\mathbf{V}_4=\begin{bmatrix}
0 & 0\\ 
1 & 0\\ 
0 & 0\\ 
0 & 1
\end{bmatrix}
\mathbf{V}_5=\begin{bmatrix}
1 & 0\\ 
0 & 0\\ 
0 & 0\\ 
0 & 1
\end{bmatrix}
\mathbf{V}_6=\begin{bmatrix}
0 & 0\\
1 & 0\\
0 & 1\\
0 & 0
\end{bmatrix}.
\end{align}
Moreover, we refer to $\{{\cal S}_j\}_{j=1}^{J}$, the set of all codebooks, by the \emph{collection of codebooks} \cite{Deka2020}, or simply the \emph{codebook collection} throughout the paper. 
A codebook collection is completely determined by the $J$ constellation matrices ${\bf C}_j, j=1,...,J$ since the mapping matrices ${\bf V}_j$ are fixed as in \eqref{mapping_matrix}. 

\subsection{SCMA Decoder}\label{Sec:SCMA_decoder}

\subsubsection{MAP Detection}

Supposing that the channel estimation is perfect and the codebook collection are available for the receiver, the detection of SCMA can be regarded as a problem of traditional multi-user detection, which can be solved by joint maximum a posteriori (MAP) \cite{SCMA_MPA}.
Then, given some received signal $\mathbf{r}$ of some block transmission (i.e., $\mathbf{r}[b]=\mathbf{r}$ for some $b$), the detected multiplexed symbol, denoted by $\widehat{\mathbf{m}}$, will be  
\begin{align}
\widehat{\bf m} = \mathrm{arg} \ \underset{{\bf m}\in {\cal Z}_M^J}{\mathrm{max}}{ \mathrm{Pr}\{\mathbf{m} ~|~{\bf r}\}}
\label{eq:all_users_MAPDetection}.
\end{align}
The $j$-th user's symbol can also be detected by maximizing its marginal a posteriori probability as follows \cite{SCMA_MPA, LDS_MPA}.
{\small
\begin{align}
\widehat{m}^{(j)}&=\mathrm{arg} \ \underset{m \in{\cal Z}_M}{\mathrm{max}}\ \sum_{\mathbf{\mathbf{m}}\in \mathcal{Z}_M^J,\ [\mathbf{m}]_j=m}\mathrm{Pr}\{\mathbf{\mathbf{m}}~|~\mathbf{r}\}
\label{eq:MAPDetection}\\
&= \mathrm{arg} \ \underset{m \in{\cal Z}_M}{\mathrm{max}}\ \sum_{\mathbf{\mathbf{m}}\in \mathcal{Z}_M^J,\ [\mathbf{m}]_j=m}\mathrm{Pr}\{\mathbf{\mathbf{m}}\}\prod_{k=1}^K f([\mathbf{r}]_k~|~\mathbf{m}).
\label{eq:product_of_functions}
\end{align}
}

Given the result of the detection above, we have $\widehat{m}^{(j)}[b]=\widehat{m}^{(j)}$ and $\widehat{\mathbf{b}}_j[b]=\widehat{\mathbf{b}}_j$ for all $j\in\mathcal{Z}_{J}$ in Fig. \ref{SystemModel} if $\mathbf{r}[b]=\mathbf{r}$.


\subsubsection{Message Passing Algorithm}
\label{SubSec:MPA}

Solving problems \eqref{eq:all_users_MAPDetection}, \eqref{eq:MAPDetection}, or \eqref{eq:product_of_functions} with brute-force has exponential complexity. Thanks to the sparsity of the codewords, the solution of this problem can be approximated by an iterative decoding algorithm, message passing algorithm (MPA), which updates the extrinsic information of function nodes (FNs) and variable nodes (VNs) along the edges in the factor graph and has moderate complexity \cite{LDS_MPA, SCMA_MPA}.


\section{Maximization of Minimum Euclidean Distance}
\label{Sec:Max_of_MED}
A codebook collection with a large MED tends to have a smaller detection error when a joint MAP detector (\ref{eq:all_users_MAPDetection}) is applied \cite{Peng2017, Lou2018} since the joint MAP detector will always choose the superimposed codeword closest to the received signal due to its highest posterior probability in \eqref{eq:all_users_MAPDetection}.
Furthermore, since the popular MPA detector usually performs very close to the joint MAP detector\cite{SCMA_ori, Zhou2017,Bao2017,Bao2019}, it makes sense to choose a codebook collection with an MED that is as large as possible.

In this section, we formulate the SCMA codebook design problem as an optimization problem that maximizes the minimum Euclidean distance (MED) under some power constraints.
We follow the definitions of MED from \cite{Peng2017, Wang2019}, that is, the MED of any two superimposed codewords of a codebook collection.

Note that in some references, MED is taken as a reasonable KPI only for the AWGN channel \cite{Chen2020}. 
However, in the scenario of 5G/NR applications \cite{3GPP_TS38211}, the considered system uses consecutive OFDM subcarriers as orthogonal resources.
It is believed that the adjacent OFDM tones tend to be nearly identical over fading channels \cite{HuaweiDownlink2014} and thus the received signal over some fading channel will be similar to the one over AWGN channel because putting identical gain on each resource is playing the same role as amplifying the noise on each resource by the reciprocal of the gain. 
Therefore, we believe that MED can be a reasonable design criterion even when frequency-selective Rayleigh fading channel is considered.



\subsection{Problem Formulation}
\subsubsection{Minimum Euclidean Distance}

For any given codebook collection determined by the matrices $\mathbf{V}_j, \mathbf{C}_j, j \in \mathcal{Z}_{J}$, we first define the square of Euclidean distance of the $k$-th and $l$-th possible superimposed codewords as
{\small\begin{align}\label{eq:d_sq} 
d_{kl} = \left\|\sum^J_{j=1} {\bf V}_j {\bf C}_j \left({\bf e}_{k_j}^{(M)} - {\bf e}_{l_j}^{(M)}\right)\right\|_2^2
\end{align}
}where $k,l\in{\cal Z}_{M^J}$ and $k_j,l_j\in \mathcal{Z}_M, \forall j\in\mathcal{Z}_J$ are defined according to the same convention as in \eqref{eq:multiplexed_symbol_index}.
Then, the minimum Euclidean distance (MED) $d_{\mathrm{min}}$ is defined as
{\small\begin{align}\label{eq:dmin} 
d_{\mathrm{min}} =\underset{\substack{ k,l\in \mathcal{Z}_{M^J}\\ k\neq l}}{\mathrm{min}} \sqrt{d_{kl}}.
\end{align}
}Note that there are totally $\binom{M^J}{2}$ possible pairs of superimposed codewords.


\subsubsection{MED Maximization Problem}\label{max_MED}

We aim to maximize the MED (\ref{eq:dmin}) subject to the power constraint. Therefore, the problem is formulated as \cite{Peng2017}

\begin{subequations}
\label{OriginProb}
\begin{align}
& \underset{\mathbf{C}\in \mathbb{C}^{N\times MJ}, t\in \mathbb{R}}{\mathrm{maximize}}
& & t \label{OriginProb1}\\
& \mathrm{subject \ to}
& & 
d_{kl}\geq t, \ \forall k,l\in \mathcal{Z}_{M^J}, k\neq l\label{OriginProb2}\\
&&&  \frac{1}{M}\mathrm{tr}(\mathbf{C}_j^H\mathbf{C}_j) = P, \forall j\in \mathcal{Z}_J \label{OriginProb3}
\end{align}
\end{subequations}
where $\mathbf{C}\triangleq [\mathbf{C}_1 \mathbf{C}_2 \cdots \mathbf{C}_{J}]$ contains the constellation matrices for all users, $t$ is an extra real-valued variable representing the square of MED, and $P$ is the limit of each user's average power of transmitted codewords. 
Here we choose equality power constraints so that each user is ensured to achieve the same power limit.
For convenience, and without loss of generality, we set $P=1$ throughout the paper.
\if01
Moreover, from a lemma to be presented later, there exists an optimal point for Problem \eqref{OriginProb} which satisfies the equality of \eqref{OriginProb3}. 
Therefore, for the convenience of applying exact penalty approach later in Section \ref{exact_penalty}, we modify \eqref{OriginProb3} into an equality constraint
\begin{equation}
\frac{1}{M}\mathrm{tr}({\bf C}_j^H{\bf C}_j) = P, ~~\forall j \in \mathcal{Z}_J \label{eq:equality_power_constraint}
\end{equation}
and call the modified problem as Problem \eqref{OriginProb}' from now on.

\begin{lemma}\label{lemma:power_equality}
If the optimal set of Problem \eqref{OriginProb} is nonempty, then there exists some optimal point such that the equality in the inequality constraint \eqref{OriginProb3} (i.e., \eqref{eq:equality_power_constraint}) holds.
\end{lemma}
The proof of Lemma \ref{lemma:power_equality} is in Appendix \ref{Lemma1_proof}.
\else
\fi

Problem \eqref{OriginProb} can be transformed into an equivalent problem in a QCQP form.
Specifically, we define $\mathbf{x} =\mathrm{vec}(\mathbf{C})\in \mathbb{C}^{n_x}$ with $n_x = NMJ$, and reformulate Problem (\ref{OriginProb}) as \cite{Peng2017}

\begin{subequations}
\label{QCQP_Prob}
\begin{align}
& \underset{\mathbf{x}\in \mathbb{C}^{n_x},t\in \mathbb{R}}{\mathrm{maximize}}
& & t \label{QCQP_Prob1}\\
& \mathrm{subject \ to}
& & \mathbf{x}^H\mathbf{A}_i\mathbf{x} \geq t , \ \forall i\in \mathcal{Z}_{\binom{M^J}{2}} \label{QCQP_Prob2} \\
&&& \mathbf{x}^H\mathbf{B}_j\mathbf{x} = MP, \ \forall j \in \mathcal{Z}_J \label{QCQP_Prob3}
\end{align}
\end{subequations}
where $\mathbf{A}_i$ and $\mathbf{B}_j$ are distance constraint and power constraint matrix, respectively. 
The matrices ${\bf A}_i$ and ${\bf B}_j$ have closed-form expressions as below. 
Note that they are real, symmetric, very sparse, and with nonzero entries limited to only values $-1$ and $1$.
%
%
The matrix ${\bf A}_i$ is
{\small
\begin{align}
{\bf A}_i &=  {\bf A}_{k, l}\nonumber\\
&= \sum_{j=1}^J\sum_{q=1}^J\left(({\bf e}^{(J)}_q)^T\otimes {\bf I}_{MN}\right)^T {\bf K}^{(q,j)}_{k,l} \left(({\bf e}^{(J)}_j)^T\otimes {\bf I}_{MN}\right)\label{eq:Ai}
\end{align}}where $${\bf K}^{(q,j)}_{k,l} = \left(({\bf e}_{k_q}^{(M)} - {\bf e}_{l_q}^{(M)})({\bf e}^{(M)}_{k_j} - {\bf e}^{(M)}_{l_j})^T\right) \otimes \left({\bf V}_q^T{\bf V}_j\right),$$
and the matrix ${\bf B}_j$ is
\begin{equation}
{\bf B}_j = \mathrm{diag}({\bf e}_j^{(J)}) \otimes {\bf I}_{NM}
\label{eq:Bj}.
\end{equation} 
The derivation of \eqref{eq:Ai} is provided in Appendix \ref{Derivation_of_A_i}.


Problem (\ref{QCQP_Prob}) is not convex, so we apply the technique of semidefinite relaxation: let $\mathbf{X} = \mathbf{x}\mathbf{x}^H \in \mathbb{H}_{+}^{n_x}$ and reformulate the problem as 
\begin{subequations} 
\label{Linear_Prob}
\begin{align}
& \underset{\mathbf{X}\in \mathbb{H}_{+}^{n_x},t\in \mathbb{R}}{\mathrm{maximize}}
& & t \label{Linear_Prob1}\\
& \mathrm{subject \ to}
& & \mathrm{tr}(\mathbf{A}_i\mathbf{X}) \geq t\ , \ \forall i\in \mathcal{Z}_{\binom{M^J}{2}} \label{Linear_Prob2} \\
&&& \mathrm{tr}(\mathbf{B}_j\mathbf{X}) = MP, \ \forall j \in \mathcal{Z}_J \label{Linear_Prob3} \\
&&& \mathrm{rank}(\mathbf{X}) = 1 \label{rank_con}
\end{align} 
\end{subequations}
Note that problems \eqref{QCQP_Prob} and \eqref{Linear_Prob} are equivalent since for any $\mathbf{X}\in\mathbb{H}^{n_x}_+$ that satisfies the rank constraint (\ref{rank_con}), there always exists some $\mathbf{x}\in\mathbb{C}^{n_x}$ (subject to a unit-norm complex ambiguity) such that $\mathbf{X} = \mathbf{x}\mathbf{x}^H$. 
\subsection{Exact Penalty Approach and Biconvex Problem Formulation}\label{exact_penalty}
Since the rank constraint \eqref{rank_con} is not a convex constraint, we can not solve it directly by the tools for solving convex optimization problems.
Therefore, we propose a method based on the concept of alternating maximization and exact penalty approach mentioned in \cite{AM, Demir2015} to obtain a rank-one solution of Problem \eqref{Linear_Prob}.
We first formulate a new problem based on Problem \eqref{Linear_Prob} as follows
\begin{subequations}
\label{biconvex_Prob}
\begin{align}
& \underset{\substack{\mathbf{X}_1,\mathbf{X}_2\in \mathbb{H}^{n_x}_{+}, \\ t_1,t_2\in \mathbb{R}}}{\mathrm{maximize}}
& & t_1 + t_2 \label{biconvex_Prob1}\\
& \mathrm{\quad subject\ to}
& & \mathrm{tr}(\mathbf{A}_i\mathbf{X}_1)\geq t_1,\ \forall i\in \mathcal{Z}_{\binom{M^J}{2}} \label{biconvex_Prob2}\\
& && \mathrm{tr}(\mathbf{A}_i\mathbf{X}_2)\geq t_2,\ \forall i\in \mathcal{Z}_{\binom{M^J}{2}} \label{biconvex_Prob3}\\
& &&\mathrm{tr}(\mathbf{B}_j\mathbf{X}_1) = MP,\ \forall j \in \mathcal{Z}_J \label{biconvex_Prob4}\\
& &&\mathrm{tr}(\mathbf{B}_j\mathbf{X}_2) = MP,\ \forall j \in \mathcal{Z}_J \label{biconvex_Prob5}\\
& && \mathrm{tr}(\mathbf{X}_1\mathbf{X}_2) = \mathrm{tr}(\mathbf{X}_1)\mathrm{tr}(\mathbf{X}_2). \label{biconvex_Prob7}
\end{align}
\end{subequations}
The following theorem shows that Problem \eqref{biconvex_Prob} is equivalent to Problem \eqref{Linear_Prob}.
\begin{theorem}
If $\{{\bf X}_1^\star, {\bf X}_2^\star, t_1^\star, t_2^\star\} $ is an optimal point for \eqref{biconvex_Prob}, then ${\bf X}_1^\star = {\bf X}_2^\star$ and $t_1^\star = t_2^\star$ and $\{{\bf X}, t\} = \{{\bf X}_1^\star, t_1^\star\}$ is an optimal point of \eqref{Linear_Prob}.
Conversely, if $\{{\bf X}^\star, t^\star\}$ is an optimal point of \eqref{Linear_Prob}, then, $\{{\bf X}_1, {\bf X}_2, t_1, t_2\} = \{{\bf X}^\star ,{\bf X}^\star, t^\star, t^\star\}$ is an optimal point of \eqref{biconvex_Prob}. 
\end{theorem}

\begin{IEEEproof}
By Theorem 1 in \cite{Demir2014}, if there are any $\mathbf{X}_1,\mathbf{X}_2\in \mathbb{H}_{+}^{n_x}$ satisfying constraint \eqref{biconvex_Prob7}, the necessary and sufficient conditions will be both of rank one and $\mathbf{X}_1=\alpha \mathbf{X}_2$ where $\alpha$ is a positive scalar. 
Suppose $\{\mathbf{X}_1^\star,\mathbf{X}_2^\star, t_1^\star, t_2^\star\}$ is an optimal point of Problem \eqref{biconvex_Prob}. 
Then, \eqref{biconvex_Prob7} implies $\mathbf{X}^\star_1=\alpha \mathbf{X}^\star_2$ for some $\alpha>0$. Since $\mathbf{X}^\star_1$ and $\mathbf{X}^\star_2$ satisfy constraints \eqref{biconvex_Prob4} and \eqref{biconvex_Prob5}, we have 
\begin{equation*}
    MP = \mathrm{tr}(\mathbf{B}_j\mathbf{X}_1^\star) = \mathrm{tr}(\mathbf{B}_j\alpha\mathbf{X}_2^\star) =  \alpha\cdot \mathrm{tr}(\mathbf{B}_j\mathbf{X}_2^\star) = \alpha MP,
\end{equation*}
which implies $\alpha = 1$ and $\mathbf{X}_1^\star = \mathbf{X}_2^\star$.
It can also be shown that $\displaystyle t_1^\star = t_2^\star = \min_{i}\mathrm{tr}({\bf A}_i{\bf X}_1^\star)$ using \eqref{biconvex_Prob2}, \eqref{biconvex_Prob3}, and \eqref{biconvex_Prob1}.
Then, we can show that $\{ {\bf X}_1^\star, t_1^\star\} $ is an optimal point of \eqref{Linear_Prob} by contradiction: if
$\{\tilde{\bf X}, \tilde{t}\}$ is some optimal point for \eqref{Linear_Prob} with $\tilde{t} > t_1^*$, then setting $\{ {\bf X}_1, {\bf X}_2, t_1, t_2\} = \{ \tilde{\bf X},\tilde{\bf X}, \tilde{t}, \tilde{t}\}$ in \eqref{biconvex_Prob} will result in a larger value in \eqref{biconvex_Prob1} ($t_1+t_2 = 2\tilde{t} > 2t_1^\star = t_1^\star + t_2^\star$).



Conversely, if $\{\mathbf{X}^\star, t^\star\}$ is an optimal point of Problem \eqref{Linear_Prob}, then  
 $\{{\bf X}_1, {\bf X}_2, t_1, t_2\} = \{{\bf X}^\star ,{\bf X}^\star, t^\star, t^\star\}$ can be shown to be an optimal point in \eqref{biconvex_Prob} by contradiction as follows.
If there is any other feasible point of Problem \eqref{biconvex_Prob}, say, $\{\mathbf{X}_1^\prime,\mathbf{X}_2^\prime, t_1^\prime, t_2^\prime\}$, such that $t_1^\prime+t_2^\prime>t_1+t_2$, then the point $\{\mathbf{X}^\prime, t^\prime\} = \{\mathbf{X}_1^\prime, t_1^\prime\}=\{\mathbf{X}_2^\prime, t_2^\prime\}$, will results in a larger value in \eqref{Linear_Prob1} ($t^\prime = (t_1^\prime+ t_2^\prime)/2 > (t_1+t_2)/2=t^\star$).
\end{IEEEproof}

To deal with constraint \eqref{biconvex_Prob7}, we apply the exact penalty approach and introduce the non-positive penalty function 
$$ \mathrm{tr}(\mathbf{X}_1\mathbf{X}_2) -\mathrm{tr}(\mathbf{X}_1)\mathrm{tr}(\mathbf{X}_2) $$ whose value is zero if and only if \eqref{biconvex_Prob7} holds \cite{Horn1991,Demir2014}.
By the equality constraints \eqref{biconvex_Prob4}, \eqref{biconvex_Prob5} and using \eqref{eq:Bj}, we have $\mathrm{tr}(\mathbf{X}_1) = \mathrm{tr}(\mathbf{X}_2) = JMP$, implying that $\mathrm{tr}(\mathbf{X}_1)\mathrm{tr}(\mathbf{X}_2)$ is a constant. Therefore, the penalty function can be further simplified as $\mathrm{tr}(\mathbf{X}_1\mathbf{X}_2)$ and then we formulate another problem as
\begin{subequations}
\label{final_Prob}
\begin{align}
& \underset{\substack{\mathbf{X}_1,\mathbf{X}_2\in \mathbb{H}^{n_x}_{+}, \\ t_1,t_2\in \mathbb{R}}}{\mathrm{maximize}}
& & t_1 + t_2 + w \cdot \mathrm{tr}(\mathbf{X}_1\mathbf{X}_2) \label{final_Prob1}\\
& \mathrm{\quad subject\ to}
& & \mathrm{tr}(\mathbf{A}_i\mathbf{X}_1)\geq t_1, \ \forall i\in \mathcal{Z}_{\binom{M^J}{2}} \label{final_Prob2}\\
& && \mathrm{tr}(\mathbf{A}_i\mathbf{X}_2)\geq t_2,\ \forall i\in \mathcal{Z}_{\binom{M^J}{2}} \label{final_Prob3}\\
& &&\mathrm{tr}(\mathbf{B}_j\mathbf{X}_1) = MP,\ \forall j \in \mathcal{Z}_J \label{final_Prob4}\\
& &&\mathrm{tr}(\mathbf{B}_j\mathbf{X}_2) = MP,\ \forall j \in \mathcal{Z}_J \label{final_Prob5}.
\end{align}
\end{subequations}
where $w > 0$ is a positive weight of the simplified penalty function $\mathrm{tr}(\mathbf{X}_1\mathbf{X}_2)$, which makes the optimization problem tend to meet constraint \eqref{biconvex_Prob7}.

Problem \eqref{final_Prob} is a biconvex optimization problem\cite{Demir2015}, meaning that it is a convex problem in ${\bf X}_1$ and $t_1$ when ${\bf X}_2$ and $t_2$ are given constants, and vice versa.
An iterative algorithm exploiting alternative maximization is presented in Algorithm \ref{biconvex}, where $\varphi_{\mathrm{max}}$ is the maximum allowable number of iterations, and, noting that the weighting $w$ can be chosen different among various iterations, $w_{1,\varphi}$ and $w_{2,\varphi}$ represent the positive weights for the $\varphi$-th iteration.

\begin{algorithm}[t]
  \caption{Proposed Algorithm based on Alternating Maximization and Exact Penalty Approach}
  \begin{algorithmic}[1]
    \Require
     $J$, $\mathbf{C}_1, \mathbf{C}_2, \cdots, \mathbf{C}_{J}$, $\mathbf{V}_1, \mathbf{V}_2, \cdots, \mathbf{V}_{J}$, $\varphi_{\mathrm{max}}$, $\{w_{1,\varphi}\}_{\varphi = 1}^{\varphi_{\mathrm{max}}}$, $\{w_{2,\varphi}\}_{\varphi = 1}^{\varphi_{\mathrm{max}}}$.
    \Ensure
    $\tilde{\mathbf{x}}$ 
    \State
    \textbf{Initialization}: Create ${\bf A}_i, \forall i \in {\cal Z}_{\binom{M^J}{2}}$ and ${\bf B}_j, \forall j \in {\cal Z}_J$ according to \eqref{eq:Ai} and \eqref{eq:Bj}.
    \State \label{step_init}
    Initialize $\mathbf{X}_2^{(0)} = \mathbf{x}\mathbf{x}^H$, where $\mathbf{x}=\mathrm{vec}([\mathbf{C}_1 \mathbf{C}_2 \cdots \mathbf{C}_{J}])$.
    \State
    Set $t_2^{(0)}=\underset{i}{\mathrm{min}}\enspace \mathrm{tr}\left(\mathbf{A}_i\mathbf{X}_2\right)$ and $\varphi =0$. 
    \Repeat
    \State   \label{step_am_1}
    Solve Problem \eqref{final_Prob} for $\{\mathbf{X}_1^{(\varphi+1)}, t_1^{(\varphi+1)}\}$ while fixing $\{\mathbf{X}_2, t_2\}$ as $\{\mathbf{X}_2^{(\varphi)}, t_2^{(\varphi)}\}$ with the weight being chosen as $w = w_{1,\varphi}$.
    \State \label{step_am_2}
    Solve Problem \eqref{final_Prob} for $\{\mathbf{X}_2^{(\varphi+1)}, t_2^{(\varphi+1)}\}$ while fixing $\{\mathbf{X}_1, t_1\}$ as $\{\mathbf{X}_1^{(\varphi+1)}, t_1^{(\varphi+1)}\}$ with the weight being chosen as $w = w_{2,\varphi}$.
    \State 
    $\varphi \leftarrow \varphi +1$
    \Until{$|\mathrm{tr}(\mathbf{X}_1^{(\varphi)})\mathrm{tr}(\mathbf{X}_2^{(\varphi)}) - \mathrm{tr}(\mathbf{X}_1^{(\varphi)}\mathbf{X}_2^{(\varphi)})|< 10^{-3}$ or $\varphi > \varphi_{\mathrm{max}}$}
    \State
    Perform singular value decomposition (SVD) on $\mathbf{X}_2^{(\varphi )}$: $\mathbf{X}_2^{(\varphi )} = \mathbf{U}\mathbf{\Sigma}\mathbf{V}^H$, with the singular values along the main diagonal of $\mathbf{\Sigma}$ in non-ascending order, i.e., $[\mathbf{\Sigma}]_{1,1} \geq [\mathbf{\Sigma}]_{2,2} \geq \cdots \geq [\mathbf{\Sigma}]_{n_x,n_x}\geq 0$.
    \State
    Set $\sigma_1=
    \begin{bmatrix}
    \mathbf{\Sigma}
    \end{bmatrix}_{1,1}$ and $\sigma_2=
    \begin{bmatrix}
    \mathbf{\Sigma}
    \end{bmatrix}_{2,2}$.
    \If{$\sigma_2 / \sigma_1 \leq 10^{-4} $}
    \State \label{step_12}{ Obtain $\tilde{\mathbf{x}}=
    \sqrt{\sigma_1}\cdot\mathbf{U}\mathbf{e}_1^{(n_x)}$}.
    \Else
    \State {Declare failure of convergence.}
    \EndIf
  \end{algorithmic}
\label{biconvex}
\end{algorithm}

Unfortunately, Algorithm \ref{biconvex} is not guaranteed to always converge within a satisfactory number of iterations. 
As will be elaborated in Section \ref{Sec:SimResult}, we observe that if the weight is kept constant during all iterations, the algorithm is more prone not to converge. 
A possible solution to this problem, accordingly to our empirical experiments, is to manually change the weights $\{w_{1,\varphi}\}, \{w_{2,\varphi}\}$.
We found that if the weights are set as a sequence that gradually increases, the algorithm tends to converge to a solution with a large MED.
\section{Dual problem of MED Maximization Problem}
\label{Sec:Dual_problem}
In this section, we derive the Lagrange dual problem associated with the primal problem \eqref{QCQP_Prob}.
Unlike the primal problem, which is non-convex, the dual problem is always a convex problem that is easier to solve.
And the optimal value of a dual problem serves as an upper bound of the optimal value of the primal problem. 
Noting that Problems \eqref{OriginProb} and \eqref{QCQP_Prob} are the two equivalent forms of the primal problem, one can derive the dual problem for each of the forms.
Here, we choose to derive the Lagrange dual of Problem \eqref{QCQP_Prob} since its QCQP form makes the dual problem derivation much easier than Problem \eqref{OriginProb}.
First of all, the Lagrangian of Problem \eqref{QCQP_Prob} is

{\small
\begin{eqnarray}
&& \mathcal{L}(\{\lambda_i\}, \{\mu_j\}, \mathbf{x},t) \nonumber\\
&=& t
+\sum\limits_{i=1}^{\binom{M^J}{2}}\lambda_{i}(\mathbf{x}^H\mathbf{A}_{i}\mathbf{x}-t)
+\sum\limits_{j=1}\limits^{J}\mu_j (MP-\mathbf{x}^H\mathbf{B}_j \mathbf{x})\nonumber\\
&=& \mathbf{x}^H\left(\sum\limits_{i=1}^{\binom{M^J}{2}}\lambda_{i}\mathbf{A}_{i}-\sum\limits_{j=1}\limits^{J}\mu_j \mathbf{B}_j\right)\mathbf{x} +\left(1-\sum\limits_{i=1}^{\binom{M^J}{2}}\lambda_{i}\right)t \nonumber\\
&& \quad  +\sum\limits_{j=1}\limits^{J}\mu_j  MP\label{Lagrangian},
\end{eqnarray}
}where we introduced Lagrange dual variables $\lambda_i$ and $\mu_j$ associated with constraints \eqref{QCQP_Prob2} and \eqref{QCQP_Prob3}, respectively.
Then, the Lagrange dual function $g(\{\lambda_{i}\},\{\mu_j\})$, defined as 
\begin{equation}
    g(\{\lambda_{i}\},\{\mu_j\}) = \sup_{\mathbf{x},t} \mathcal{L}(\{\lambda_i\}, \{\mu_j\}, \mathbf{x},t),
\end{equation}
is unbounded above if any eigenvalue of
$\sum_{i=1}^{\binom{M^J}{2}}\lambda_{i}\mathbf{A}_{i}-\sum_{j=1}^{J}\mu_j \mathbf{B}_j
$ is greater than zero or $\sum_{i=1}^{\binom{M^J}{2}}\lambda_{i} \neq 1$.
Otherwise, the dual function is 
\begin{equation}
g(\{\lambda_{i}\},\{\mu_j\}) = \sum\limits_{j=1}\limits^{J}\mu_j MP.
\end{equation}
Therefore, the Lagrange dual problem is found to be
\begin{subequations} 
\label{Lagrangian_dual_problem}
\begin{align}
& \underset{\{\lambda_{i}\},\{\mu_j\}}{\mathrm{minimize}}
& &  \sum\limits_{j=1}\limits^{J}\mu_j MP \label{Lagrangian_dual_problem1}\\
& \mathrm{subject \ to}
& & 
	\sum\limits_{i=1}^{\binom{M^J}{2}}
	\lambda_{i}\mathbf{A}_{i}
	\preceq\sum\limits_{j=1}\limits^{J}\mu_j \mathbf{B}_j
 \label{Lagrangian_dual_problem2} \\
&&& \sum\limits_{i=1}^{\binom{M^J}{2}}\lambda_{i}= 1 \label{Lagrangian_dual_problem3}\\
&&& \lambda_{i}\geq 0, \ \forall i \in \mathcal{Z}_{\binom{M^J}{2}}. \label{Lagrangian_dual_problem5}
\end{align}
\end{subequations}

It is well known that weak duality \cite{ConvexOptimization} dictates that the optimal value of the dual problem \eqref{Lagrangian_dual_problem} is an upper bound of the optimal value of the primal problem \eqref{QCQP_Prob}, as also that of 
\eqref{OriginProb} since problems \eqref{OriginProb} and \eqref{QCQP_Prob} are equivalent. 
In fact, as we will find later in Section \ref{Sec:SimResult}, the optimal values of primal problem \eqref{QCQP_Prob} and dual problem \eqref{Lagrangian_dual_problem} will coincide, at least for the case of $J=3$, suggesting that strong duality holds for this case.
\section{Simulation Results}
\label{Sec:SimResult}

In this section, we conduct numerical simulations to verify the proposed methods presented in Section \ref{Sec:Max_of_MED} and compare their performances with existing methods \cite{starQAM2015, Huawei2015, Zhang2016, Chen2020, Deka2020}, and also the MED upper bound derived in Section \ref{Sec:Dual_problem}. 
Throughout all simulations, we set the number of resources as $K=4$, the cardinality of codebooks as $M=4$, the constellation sizes as $N=2$, and 15 times message passing iterations. 
The definitions of SER and BER are shown as follows.
\begin{definition}
The symbol error rate (SER) is defined as
{\small
\begin{align}
    P_{e,s} &=  \frac{1}{J}\sum_{j=1}^{J} \sum_{m=1}^{M}\mathrm{Pr}\{ m^{(j)}=m\} \mathrm{Pr}\{\widehat{m}^{(j)}\neq m \enspace |\enspace m^{(j)}=m\} \label{SER_def}
\end{align}
}where, for all $j\in \mathcal{Z}_{J}$, $m^{(j)}$ and $\widehat{m}^{(j)}$ follow the definitions in Section \ref{Sec:SMPM}
, and $ \mathrm{Pr}\{m^{(j)}=m\}$ is assumed to be $1/M$ for all $m\in \mathcal{Z}_{M}$.
\end{definition}
\begin{definition}
The bit error rate (BER) is defined as
{\small
\begin{align}
    P_{e,b} 
    &=  \frac{1}{J\mathrm{log}_2 M}\sum_{j=1}^{J} \sum_{i=1}^{\mathrm{log}_2 M}\Big( \mathrm{Pr}\{[\mathbf{b}_j]_i = 0\} \mathrm{Pr}\{[\widehat{\mathbf{b}}_j]_i=1~|~[\mathbf{b}_j]_i=0\} \nonumber\\
    & \qquad +\mathrm{Pr}\{[\mathbf{b}_j]_i = 1\} \mathrm{Pr}\{[\widehat{\mathbf{b}}_j]_i=0~|~[\mathbf{b}_j]_i=1\}\Big)
\end{align}
}where, for all $j\in \mathcal{Z}_{J}$, $\mathbf{b}_j$, $\widehat{\mathbf{b}}_j$ follow the definitions in Section \ref{Sec:SMPM}
, and $ \mathrm{Pr}\{[\mathbf{b}_j]_i=0\}$ and $ \mathrm{Pr}\{[\mathbf{b}_j]_i=1\}$ are assumed to be $1/2$ for all $i\in \mathcal{Z}_{\mathrm{log}_2 M}$.
\end{definition}
Notations $E_{s}$ and $E_{b}$, representing the average energy of all users' symbols and data bits, respectively, are defined as
\begin{align} \label{E_s}
    E_{s} &= 
    \frac{1}{JM} \sum\limits_{j=1}^{J}\sum\limits_{m=1}^{M}\left \|\mathbf{V}_j\mathbf{C}_j\mathbf{e}_m^{(M)} \right \|_2^2 =\frac{1}{JM}\sum\limits_{j=1}^{J}\mathrm{tr}({\bf C}_j^H{\bf C}_j)
\end{align}
and
\begin{align}
    E_b &= \frac{1}{\mathrm{log}_2 M} E_s.
\end{align}
The normalized minimum Euclidean distance is defined as
\begin{equation}
    \hat{d}_{\mathrm{min}} = \frac{d_{\mathrm{min}}}{\sqrt{E_s}}
\end{equation}
where 
$E_s$ is also the average power of each codeword of each user by the definition in \eqref{E_s}.
\subsection{The Case with Three Users ($J=3$)}
We first consider the simple case where only three users are allowed to share the $K=4$ orthogonal resources, i.e., $J=3$. 
It will be demonstrated that, in this simple case, it is possible to use Algorithm \ref{biconvex} to obtain a codebook collection whose MED reaches the theoretical upper bound determined by \eqref{Lagrangian_dual_problem}.
Mapping matrices $\mathbf{V}_1$, $\mathbf{V}_2$, $\mathbf{V}_3$ in \eqref{mapping_matrix} are chosen and this choice makes sure the number of collisions is as small as 2 \cite{Peng2017}.
Algorithm \ref{biconvex} is applied to solve Problem \eqref{final_Prob} with the initial point $\mathbf{X}_2=\mathbf{x}\mathbf{x}^H$ where all elements of $\mathbf{x}$ (i.e., all entries of $\{{\bf C}_j\}_{j=1}^3$) are independently generated by a random variable uniformly distributed between 0 and 1.
In this relatively simple case, Algorithm \ref{biconvex} is found to be converging and leading to an optimal codebook collection whose optimal value coincides with that of the dual problem (\ref{Lagrangian_dual_problem}).
We used \texttt{CVX}, a MATLAB-based modeling system for convex optimization \cite{cvx}, for Steps \ref{step_am_1} and \ref{step_am_2} of Algorithm \ref{biconvex} in each iteration. 
As a result, we observed that, in about 10 iterations, the algorithm converged. 
We also used \texttt{CVX} to solve the dual problem (\ref{Lagrangian_dual_problem}) to get the upper bound of optimal value.
Note that the dual problem \eqref{Lagrangian_dual_problem} is always convex, so it can be solved with just a single \texttt{CVX} instance.

\begin{figure}[ht]
\centering \centerline{
\includegraphics[width=0.54\textwidth,clip]{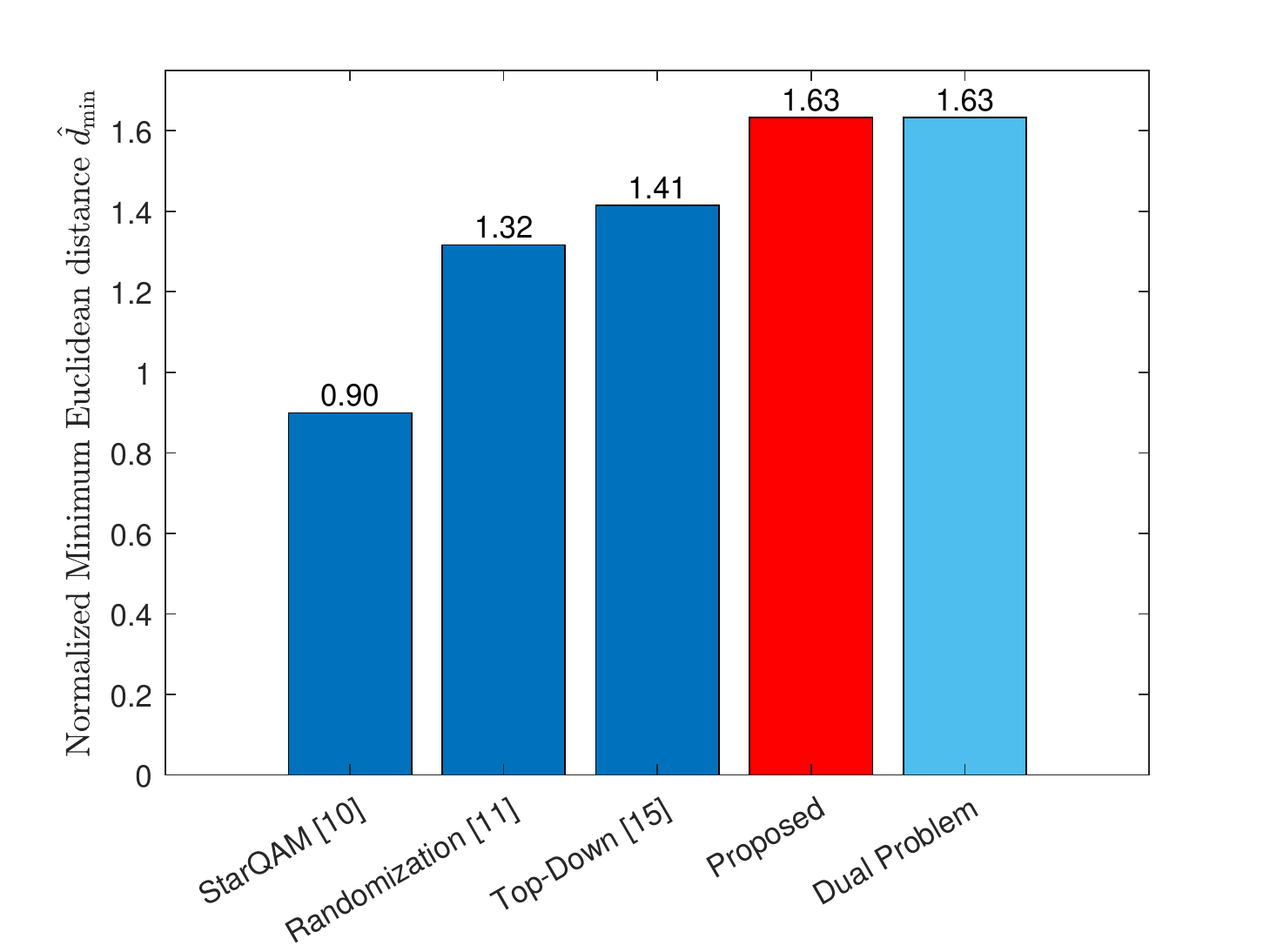}}
\caption{Minimum Euclidean distance comparison for $J=3$.}
\label{fig:dmin_comp_J3}
\end{figure}

Fig. \ref{fig:dmin_comp_J3} shows the MED comparison of the proposed algorithms and various previously reported methods in terms of MED, along with the bound given by the dual problem (\ref{Lagrangian_dual_problem}). 
As indicated in the figure, the proposed method achieves the largest value among all methods, including the randomization method \cite{Peng2017} with $L_{rand}=10^6$ \footnote{The codebook collection obtained here do not necessarily coincide with the one reported in \cite{Peng2017}, due to the random nature of Gaussian randomization algorithm. }, the starQAM codebook collection \cite{starQAM2015}, and the Top-Down codebook collection \cite{topdown2016}. 
We notice that the MED of the proposed method is the same as the optimal value of the dual problem \eqref{Lagrangian_dual_problem}. 
As stated in Section \ref{Sec:Dual_problem}, it is  sufficient to show that the codebook collection we proposed is a set of optimal codebooks that achieve the maximum MED. 
\subsection{The Case with Six Users ($J=6$)}


We now consider the case when the number of users is $J=6$. As mentioned in Section \ref{Sec:Max_of_MED}, a large number of distance constraints greatly increase the consumption of memory and computational complexity. 
It is extremely inefficient for the \texttt{CVX} tool to handle a large number of more than 8 million constraints in constructing the problem settings alone, not to mention solving it.
As an alternative approach, we resort to directly using SDPT3 \cite{sdpt3}, the default solver of the \texttt{CVX} tool.
We observed that the SDPT3 solver, without the extra burden caused by the problem-constructing routines of \texttt{CVX}, is capable of returning correct results within an acceptable time duration.
Moreover, it is worthy to note that the sparse properties of the matrices $\mathbf{A}_i, \mathbf{B}_j$ with nonzero entries being $\pm 1$, may have also expedited the computation of the SDPT3 solver \cite{sdpt3}. 
We used this new approach to execute Algorithm \ref{biconvex} and tried to solve Problem \eqref{QCQP_Prob}. 
Although it may not always converge within $\varphi_{\mathrm{max}}$ iterations, we usually can obtain an $\mathbf{X}$ with the largest eigenvalue dominating all the other eigenvalues (i.e., 
the ratio of the second largest eigenvalue to the largest one $\sigma_2/\sigma_1 \leq 10^{-4}$) and thus it is fair enough to consider $\mathbf{X}$ as $\mathbf{x}\mathbf{x}^H$ for retrieving $\mathbf{x}$ by Step \ref{step_12} in Algorithm \ref{biconvex}.

Since Algorithm \ref{biconvex} needs the constellation matrices $\{{\bf C}_j\}_{j=1}^6$ of some codebook collection for initialization, we chose two codebook collections for AWGN channel proposed in \cite{Chen2020} and \cite{Deka2020} since they have relatively good error rate performances comparing to other codebook collections over AWGN channels.
We first test Algorithm \ref{biconvex} using the codebook collection for AWGN channel proposed in \cite{Chen2020} (referred to as "Chen's AWGN codebook collection") for initialization, and we chose weight $w=0.1$.
The algorithm converged in $18$ iterations with a total computation time of $26$ hours.
The MED of the resultant codebook collection is 1.17, which is already greater than the MEDs of all previously reported codebook collections (The detailed codebook collection is shown in Appendix \ref{Chen_init}).
Then, we tried Algorithm \ref{biconvex} using the codebook collection for AWGN channel proposed in \cite{Deka2020} (referred to as "Deka's AWGN codebook collection")\footnote{
Note that the "Chen's AWGN codebook collection" \cite{Chen2020} and "Deka's AWGN codebook collection" \cite{Deka2020} for initialization we used here are scaled to meet power constraint \eqref{OriginProb3}.}.
We at first tried the fixed weight $w=0.3$ but it ran for over $80$ iterations, which takes almost a whole week of computing with the CPU being AMD Ryzen™ 9 3900X, and did not converge. 
Hoping for the convergence of Algorithm \ref{biconvex}, we manually chose weight $w$ between $0.1$ and $0.35$ in the process of the iterations in Algorithm \ref{biconvex} and it converged with MED being 1.30 (The detailed codebook collection is shown in Appendix \ref{Deka_init}). 
More specifically, we chose:

{\footnotesize
\begin{equation*}
    w_{1,\varphi} = \left\{\begin{array}{ll}
    0.1     &  1 \leq \varphi \leq  6\\
    0.15    &  7 \leq \varphi \leq 27\\
    0.2     & 28 \leq \varphi \leq 35\\
    0.25    & 36 \leq \varphi \leq 59\\
    0.3     & 60 \leq \varphi \leq 71\\
    0.35    & 72
    \end{array}\right.,
    w_{2,\varphi} = \left\{\begin{array}{ll}
    0.1     &  1 \leq \varphi \leq  6\\
    0.15    &  7 \leq \varphi \leq 26\\
    0.2     & 27 \leq \varphi \leq 34\\
    0.25    & 35 \leq \varphi \leq 58\\
    0.3     & 59 \leq \varphi \leq 71\\
    0.35    & 72
    \end{array}\right..
\end{equation*}
}It took $72$ iterations and $114$ hours to obtain this codebook collection.

\begin{figure}
    \centering \centerline{
    \includegraphics[width=0.54\textwidth,clip]{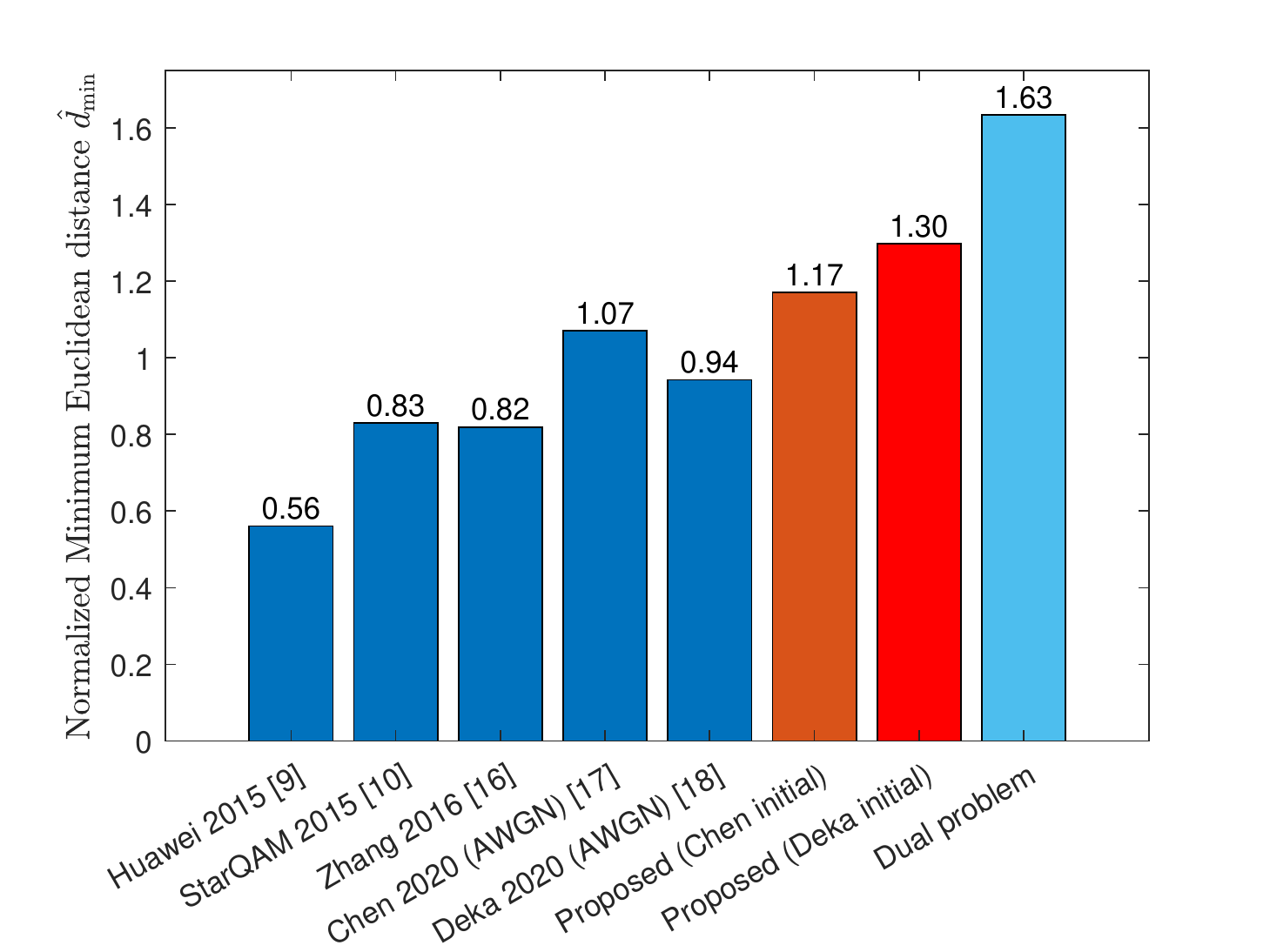}}
    \caption{Minimum Euclidean distance comparison for $J=6$.}
    \label{fig:dmin_comp_J6}
\end{figure}

The dual problem \eqref{Lagrangian_dual_problem} is also solved via SDPT3 with $J=6$ and the resulting upper bound is 1.63.
The comparison of the normalized MEDs of different codebook collections is shown in Fig. \ref{fig:dmin_comp_J6} \footnote{
Note that all codebook collections we used for comparison hereafter are scaled such that the power of the user with maximum average power meet power constraint \eqref{OriginProb3} (i.e., $\underset{j\in\mathcal{Z}_{J}}{\mathrm{max}}\ \frac{1}{M} \mathrm{tr}\left({\mathbf{C}_j^H\mathbf{C}_j}\right)=P$ for all codebook collections).}. 
Although Deka's AWGN codebook collection (labeled as ``Deka 2020 (AWGN) \cite{Deka2020}'') have a smaller MED than Chen's AWGN codebook collection (labeled as ``Chen 2020 (AWGN) \cite{Chen2020}''), the former results in a codebook collection with an even larger MED. 
We found that this may have been due to that only a very small number of pairs of superimposed codewords achieve the MED for Deka's AWGN codebook collection. 

As the codebook collection obtained by Algorithm \ref{biconvex} with Deka's AWGN codebook collection \cite{Deka2020} for initialization has the largest MED, 1.30, it is expected that it will attain relatively better error rate performances than other codebook collections and therefore we use it as the proposed codebook collection in the following numerical results. 

The parameter setting is shown as follows. The FFT size $N_B$ is 256.
The channel length $L$ is 18 and cyclic-prefix length $N_{CP}$ is 17. 
The subcarrier indices of the $K$-dimensional SCMA signal is set to 127, 128, 129, 130 within the 256 subcarriers of an OFDM symbol (i.e. $i_{sub}=127$ in \eqref{received_signal}). 
The other signals loaded on the remaining subcarriers are set as independent and identically distributed (i.i.d.) signals with distribution being $\mathcal{CN}(0,JE_s/K)$.
We adopted the MPA algorithm described in Section III-B in reference \cite{SCMA_MPA}  for the detection. 
The results, corresponding to the AWGN channel and Rayleigh fading channel, are discussed in the following parts respectively.

\subsubsection{AWGN Channel}
For the simulation of the downlink SCMA system based on OFDMA over AWGN channel, the results, as shown in Fig.  \ref{fig:SCMA_Codebook_comparision_J6_SER}, and  \ref{fig:SCMA_Codebook_comparision_J6_BER}, demonstrate that the proposed codebook collection obtained by Algorithm \ref{biconvex} with Deka's AWGN codebook collection \cite{Deka2020} for initialization indeed achieves the best SER and BER performances since it has the largest MED.
Specifically, there are gains of both about 0.7dB at SER $= 10^{-5}$ and BER $= 10^{-5}$ over the best existing codebook collection \cite{Deka2020}.

Moreover, since the downlink SCMA system based on OFDMA over AWGN channel has an all-one channel gain for all subchannels, and the additive noises after the FFT operation are still uncorrelated Gaussian noises with zero means and $N_0$ variances, this case is actually the same case with AWGN channel considered in other works \cite{starQAM2015, Peng2017, Alam2017, Zhou2017, Sharma2018, topdown2016, Zhang2016, Chen2020, Deka2020}.

\subsubsection{Rayleigh Fading Channel}
\begin{figure}[ht]
    \centering\centerline{
    \includegraphics[width=0.54\textwidth,clip]{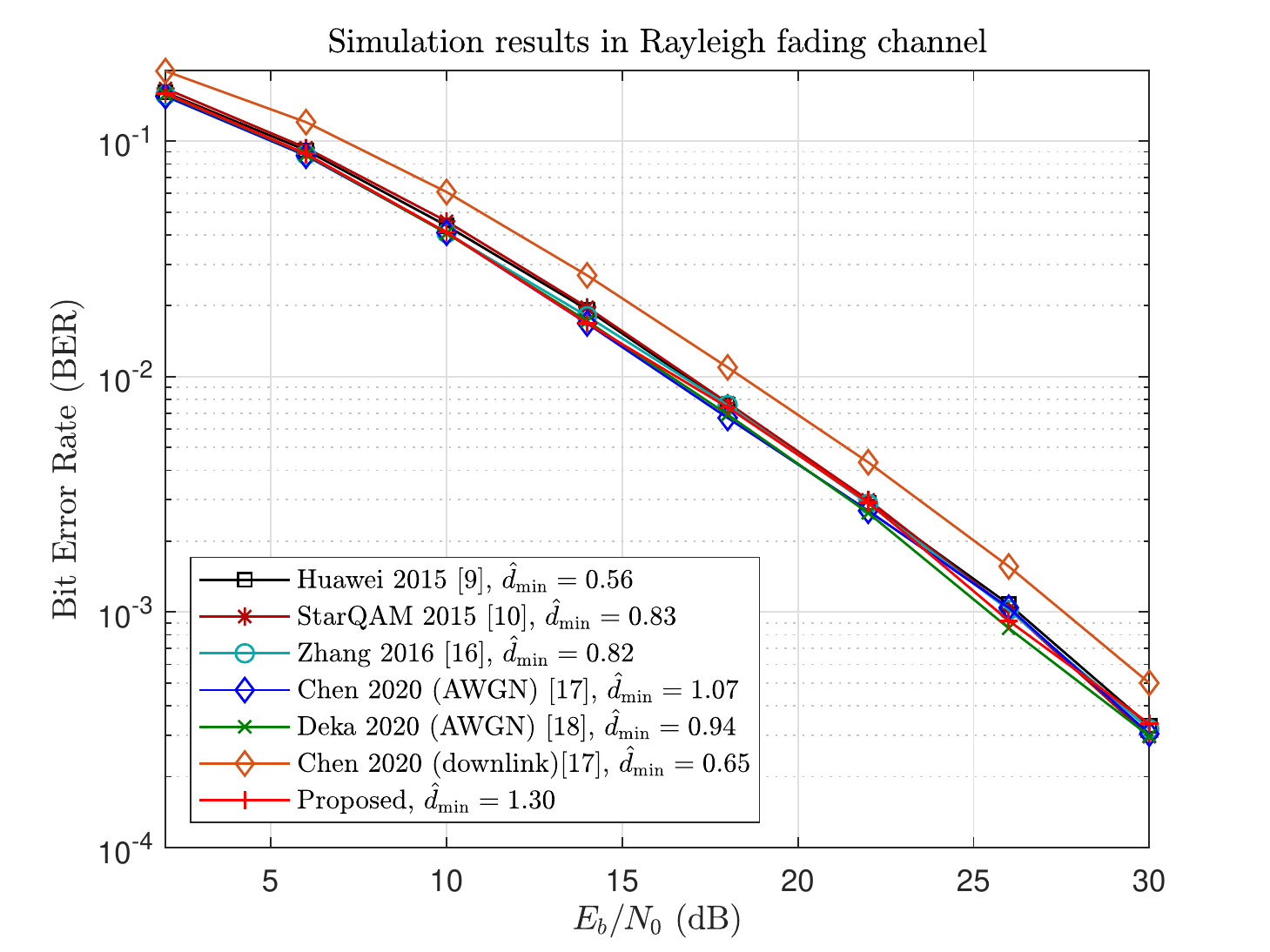}}
    \caption{BER performance comparison for $J=6$ over Rayleigh fading channel (including cases with poor channel).}
    \label{fig:BER_simulation_fading}
\end{figure}

The Rayleigh fading channel here is set to be a frequency-selective channel, which fits in the scenario with multipath channel in 5G/NR applications (e.g. TDL-A/C, EPA) \cite{3GPP_38802} and is, for convenience, specifically set as
\begin{equation}
\label{eq:fading_channel}
    \mathbf{h}\sim \mathcal{CN}(0,\mathrm{diag}(\sigma_{h[0]}^2, \sigma_{h[1]}^2,\dots, \sigma_{h[L-1]}^2 ))
\end{equation}
where $\left[\sigma_{h[0]}~\sigma_{h[1]}~\cdots~ \sigma_{h[L-1]}\right]^T=\mathbf{h}_{\sigma}/\left \|\mathbf{h}_{\sigma}\right \|_2$, and $\mathbf{h}_{\sigma}\in \mathbb{R}^{L}$ is the vector whose elements are linearly spaced between $0$ dB and $-48$ dB.
With the setting above, the simulation results are shown in Fig. \ref{fig:BER_simulation_fading}. We notice that most of the codebook collections have similar performances on bit error rate, but the codebook collection proposed by Chen \textit{et al.} \cite{Chen2020} for downlink Rayleigh fading channel (labeled as ``Chen 2020 (downlink) \cite{Chen2020}'') has worse performances. 

Although the proposed codebook collection does not outperform other ones in BER for the case mentioned above, we notice that $E_b/N_0$ needs to be about 26dB for BER$=10^{-3}$, and thus $E_b/N_0$ is expected to be unrealistically high for a better BER, such as $10^{-5}$ or $10^{-6}$.
The bad performance curves for all methods may have been mostly contributed by deep fade channels in some Monte Carlo trials.
In the 5G/NR applications, however, we believe it is reasonable to assume all users that share the SCMA resources would possess sufficiently good channel quality on these subcarriers since the base station, with the knowledge of channel state information reported from a user (e.g., see Section 5 of \cite{3GPP_TS38214}), is likely to assign users to the resources with good channel quality.
With this assumption in mind, we conduct the performance analysis again by excluding the $40$ percent poorest channels $\mathbf{h}^f_{sub}$.
The results, as shown in Fig. \ref{fig:BER_simulation_fading_dicardBad}, demonstrate that the proposed codebook collection outperforms all the other codebook collections on bit error rate. Specifically, there is a gain of about $0.6$ dB at BER $= 10^{-5}$ over the best existing codebook collection \cite{Deka2020}.

\subsection{Simulation Results under Other Channel Models}\label{Sec:other_cases}
In this subsection, we conduct simulations under some channel models that have been considered in \cite{Mheich2019, Deka2020, Chen2020}, whose system models are slightly different from the one we described in Section \ref{downlink_SCMA_system_based_on_OFDMA}. 
Specifically, they consider the downlink and uplink SCMA systems over OFDMA with separate subcarriers.
Even though the 5G/NR standards have adopted resource blocks that are composed of consecutive subcarriers rather than separate ones \cite{3GPP_TS38211}, we investigate these two cases for a more comprehensive comparison of the proposed codebook collection obtained in the previous subsection (as shown in Appendix \ref{Deka_init}) with existing ones.
\subsubsection{Downlink Rayleigh Fading Channel with Non-Consecutive Subcarriers}
\begin{figure}[ht]
    \centering\centerline{
    \includegraphics[width=0.54\textwidth,clip]{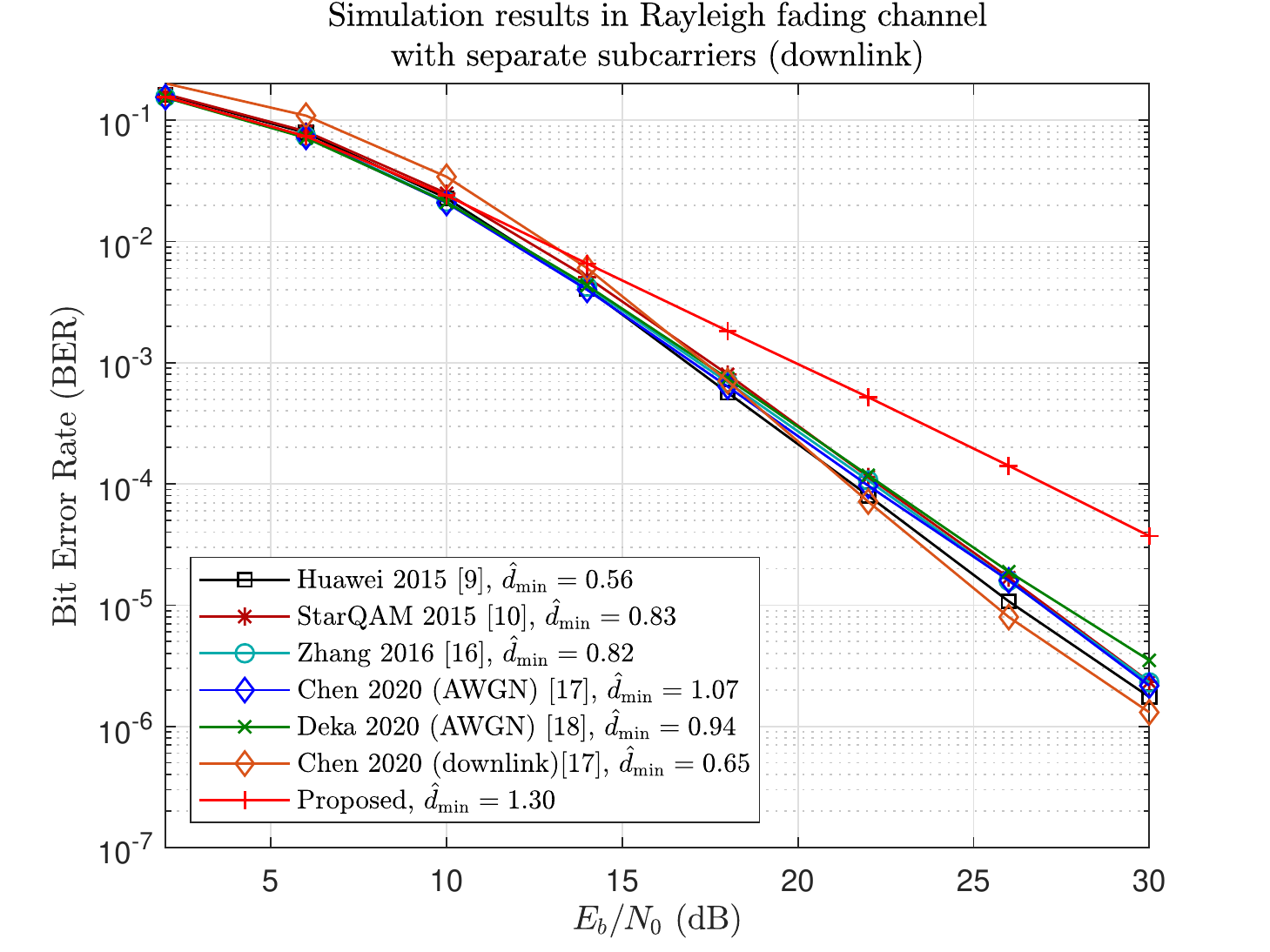}}
    \caption{BER performance comparison for $J=6$ over Rayleigh fading channel with separate subcarriers (downlink).}
    \label{fig:BER_simulation_downlink_separate}
\end{figure}

In \cite{Chen2020}, the authors considered this type of channel as an equivalent SISO fast Rayleigh channel since the statistics of channels of non-adjacent subcarriers separated by a frequency gap greater than the coherence bandwidth could be considered to be uncorrelated.
To simulate this scenario, we slightly modified the system model depicted in Fig. \ref{SystemModel} and make ${\bf s}_1[b]$ to ${\bf s}_J[b]$ occupy non-consecutive subcarriers, so the received signals at the corresponding subcarriers can be expressed as 
\begin{equation}
{\bf r}[b] = \mathrm{diag}({\bf h}_{\mathrm{sub, sep}}^f) {\bf s}[b] + {\bf n}[b]
\end{equation}
where ${\bf h}_{\mathrm{sub, sep}}^f = [[{\bf h}^f]_{i_1}~[{\bf h}^f]_{i_2}~\cdots[{\bf h}^f]_{i_K}]^T\in\mathbb{C}^K$ is the subvector of the frequency-domain channel gain vector ${\bf h}^f \in \mathbb{C}^{N_B}$ at subcarriers indexes $i_1$ through $i_K$ that are non-consecutive or even separated apart.
We choose $(i_1, i_2, i_3, i_4) = (32, 96, 160, 224)$ in our simulation and the results are shown in Fig. \ref{fig:BER_simulation_downlink_separate}.
The results match the corresponding results shown in \cite{Chen2020}\footnote{Note that the shift in $E_b/N_0$ between our results and the corresponding ones in \cite{Chen2020} is contributed by the different setting of $E_s$.}, reconfirming the equivalence of the SISO fast Rayleigh channel considered therein and the Rayleigh fading channel using widely separated subcarriers under OFDMA. 
We observe that the proposed codebook collection, compared to others, does not work well perhaps because the product distance \cite{Chen2020} of the codebook collection is not optimized.

\subsubsection{Uplink Rayleigh Fading Channel}
\begin{figure}[ht]
    \centering\centerline{
    \includegraphics[width=0.54\textwidth,clip]{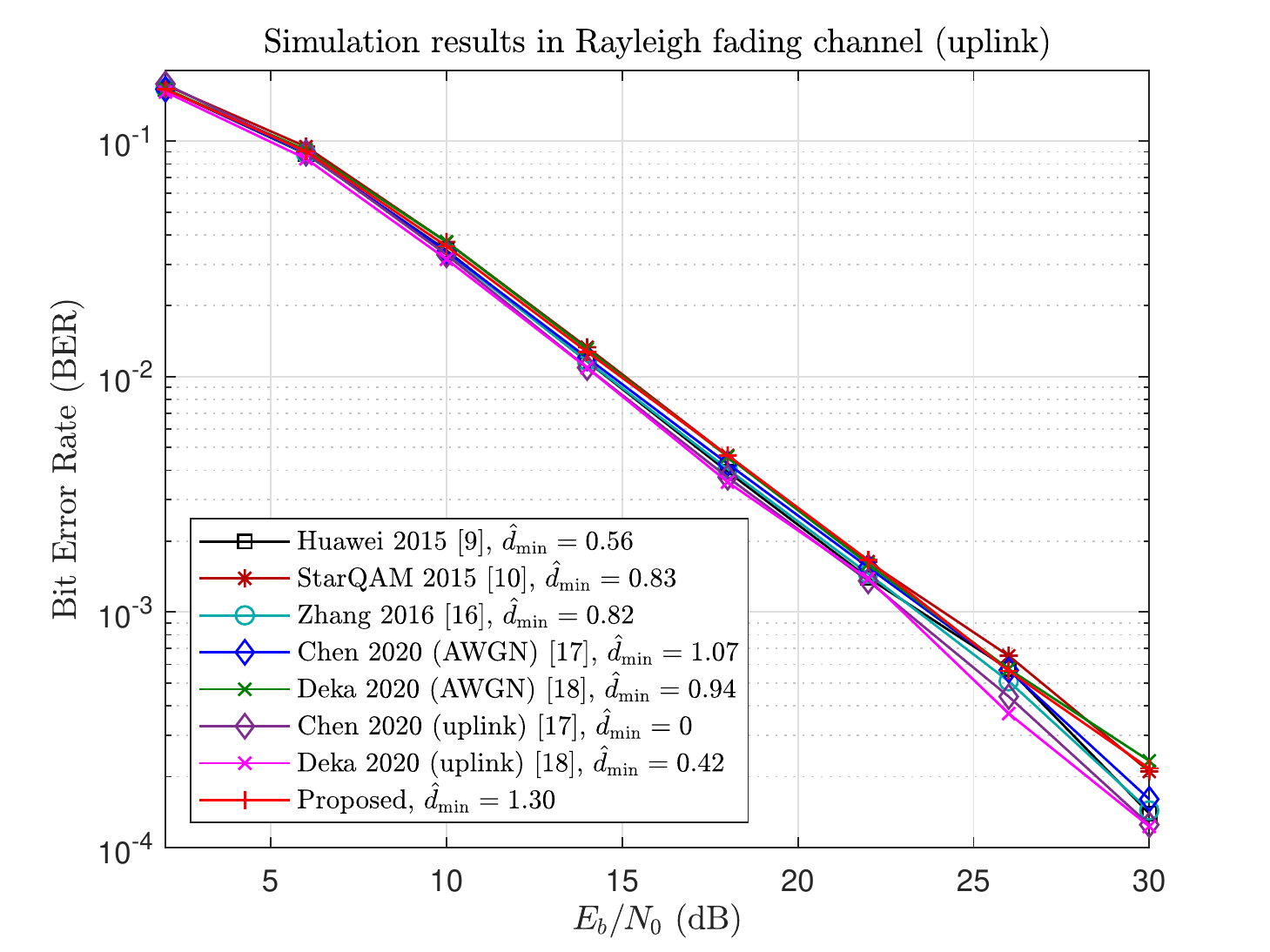}}
    \caption{BER performance comparison for $J=6$ over Rayleigh fading channel with consecutive subcarriers (uplink).}
    \label{fig:BER_simulation_uplink}
\end{figure}

For the uplink SCMA system based on OFDMA, it can be regarded as that the transmitter of each user is transmitting the OFDM signals with its SCMA signals, say, $\mathbf{s}_j[b]$, loaded on the chosen subcarriers, and the receiver receives the $J$ synchronized signals from the $J$ transmitters and process the received signal in the same way as in the downlink SCMA system based on OFDMA.
We first consider the case with consecutive subcarriers and the corresponding received signal can be expressed as
\begin{equation}
\mathbf{r}[b] = \sum_{j=1}^{J}\mathrm{diag}(\mathbf{h}_{j,sub}^f)\mathbf{s}_j[b]+\mathbf{n}[b], 
\end{equation}
where $\mathbf{h}_{j,sub}^f=\left[[\mathbf{h}^f_j]_{i_{sub}}~ [\mathbf{h}^f_j]_{i_{sub}+1}~ \cdots ~[\mathbf{h}^f_j]_{i_{sub}+K-1}\right]^T\in \mathbb{C}^{K}$ is the subvector of $\mathbf{h}^f_j$, the frequency-domain channel gain vector for the $j$-th user.
Vector $\mathbf{h}^f_j$ is the discrete Fourier transform of vector $\left[\mathbf{h}_j^T \enspace \mathbf{0}^T_{N_B-L} \right]^T\in \mathbb{C}^{N_B}$, where $\mathbf{h}_j$ shares the same distribution as the $\mathbf{h}$ in \eqref{eq:fading_channel}.
We assume that $\mathbf{h}_{j_1}^f$ and $\mathbf{h}_{j_2}^f$ are statistically uncorrelated whenever $j_1\neq j_2$.
With the setting above, the simulation results are shown in Fig. \ref{fig:BER_simulation_uplink}\footnote{The codebook collections for uplink systems proposed in \cite{Chen2020, Deka2020} are labeled as ``Chen 2020 (uplink) \cite{Chen2020}'', ``Deka 2020 (uplink) \cite{Deka2020}'', respectively.}. 
The proposed codebook collection does not outperform other ones.
For the uplink case with non-consecutive subcarriers, as the readers can imagine, the proposed codebook collection will have an even worse performance.
The relatively bad BER performances of the proposed codebook collection indicate the fact that MED may not be a sufficient KPI for the uplink case, which reconfirms the claim of weak correlation of MED and BER performances in the uplink case in \cite{Chen2020}.

\begin{figure}[ht]
    \centering\centerline{
    \includegraphics[width=0.54\textwidth,clip]{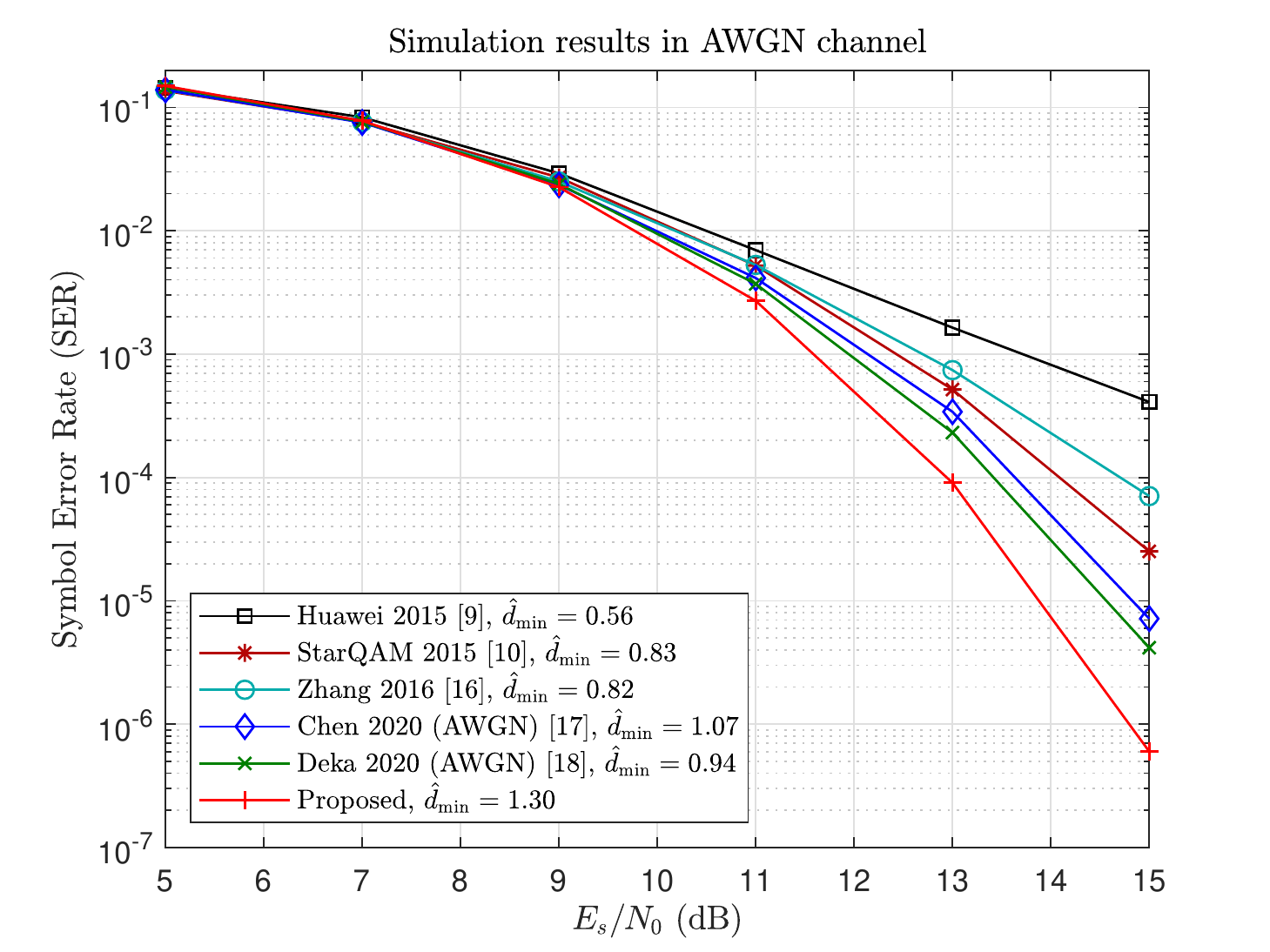}}
    \caption{SER performance comparison for $J=6$ over AWGN channel.}
    \label{fig:SCMA_Codebook_comparision_J6_SER}
\end{figure}

\begin{figure}[ht]
    \centering\centerline{
    \includegraphics[width=0.54\textwidth,clip]{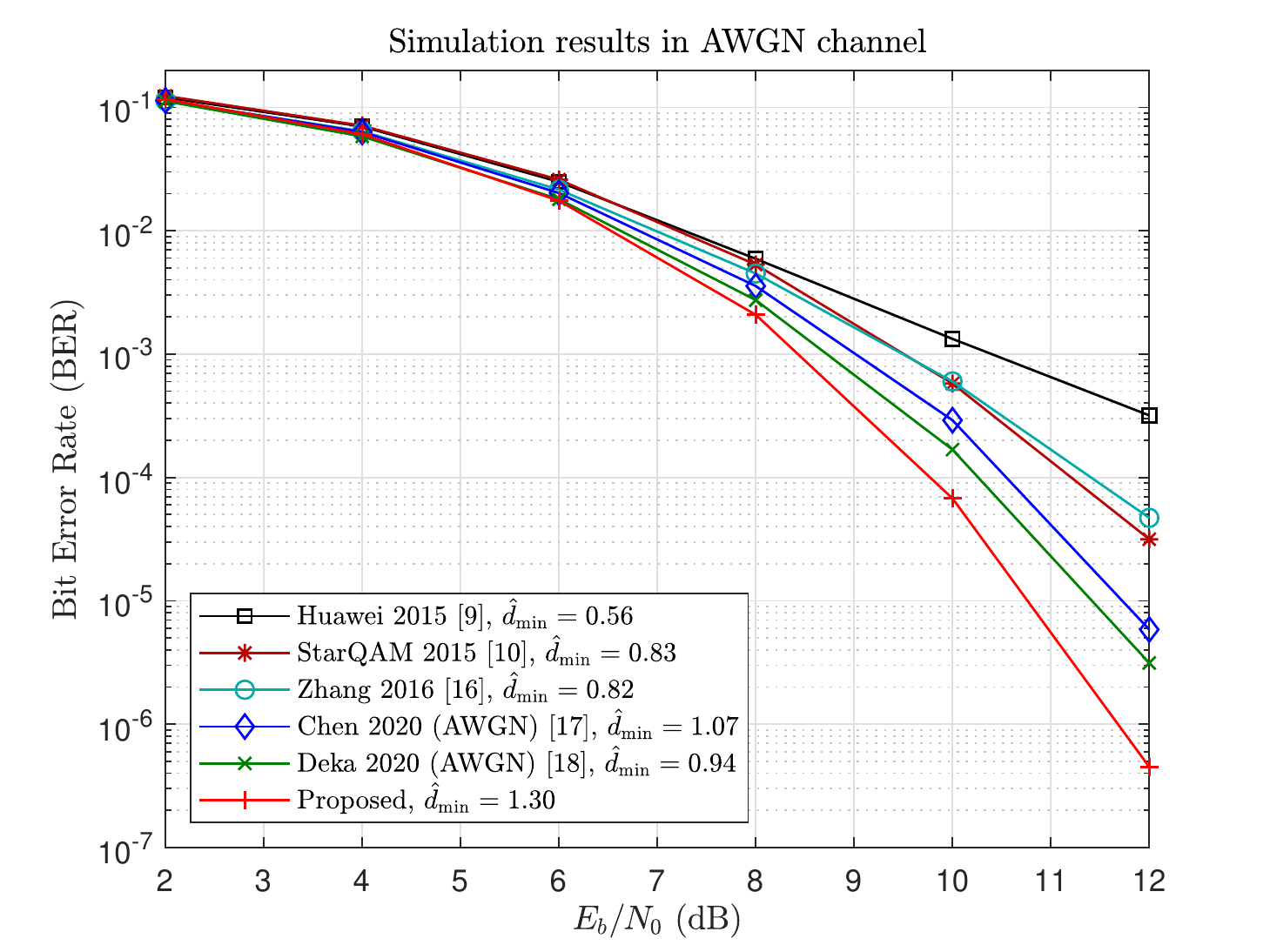}}
    \caption{BER performance comparison for $J=6$ over AWGN channel.}
    \label{fig:SCMA_Codebook_comparision_J6_BER}
\end{figure}

\begin{figure}[ht]
    \centering\centerline{
    \includegraphics[width=0.54\textwidth,clip]{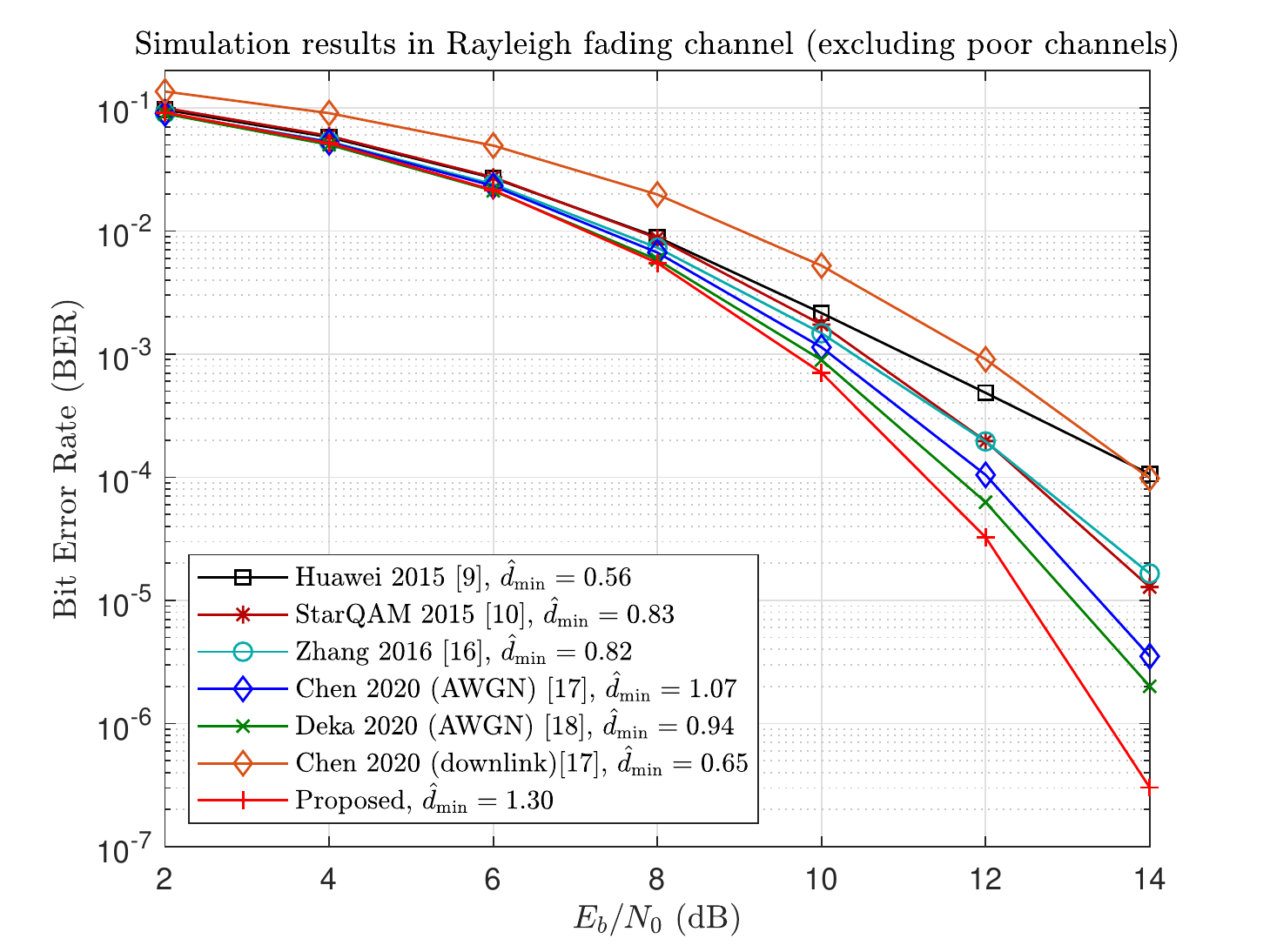}}
    \caption{BER performance comparison for $J=6$ over Rayleigh fading channel (excluding cases with poor channel).}
    \label{fig:BER_simulation_fading_dicardBad}
\end{figure}


\subsection{Comparison of Theoretical Results and Simulation Results}
\label{Sec:formula_sim_Comparison}
To certify the error rate performances reported in the previous simulation plots, we compare all error rate curves of the cases in AWGN channel with theoretical bounds in this subsection. 
The theoretical bounds are derived based on the maximum likelihood (ML) multi-user detection \cite{Bao2017, Bao2019}, which is equivalent to joint MAP detection due to the equally likely input symbols. 
Moreover, because of the similar error rate performance in MPA detection and joint MAP detection \cite{SCMA_ori, Zhou2017,Bao2017, Bao2019}, it is reasonable to compare the MPA-based simulation results and the bounds of the error rates of the MAP detector. Hence, we present these bounds in the following paragraph for the comparison with the simulation results.


For the SER of the MAP detector, denoted by $P_{e,s}^{\mathrm{MAP}}$, an upper bound can be found by simply taking the average over $J$ users of the upper bound of the SER of each single user derived by Bao \textit{et al}. (Eq. (38) in \cite{Bao2017}).
\begin{equation}
\label{SER_formula}
    P_{e,s} \approx P_{e,s}^{\mathrm{MAP}} \leq \frac{1}{M^J\cdot J}\mathbf{1}^T_{M^J} (\mathbf{Q}\circ \mathbf{D}_s )\mathbf{1}_{M^J} 
\end{equation}
where the matrix $\mathbf{Q}\in \mathbb{R}^{M^J\times M^J}$ is defined with $[\mathbf{Q}]_{k,l}=Q\left(\sqrt{\frac{d_{kl}}{2N_0}}\right)$, $d_{kl}$ was defined in \eqref{eq:d_sq}, the matrix $\mathbf{D}_s\in \mathbb{Z}^{M^J\times M^J}$ is defined with $[\mathbf{D}_s]_{k,l} = d_{s,kl}$,  
and $d_{s,kl}$ is the Hamming distance between the multiplexed symbols $\mathbf{m}_k, \mathbf{m}_l$, i.e., $ d_{s,kl} = \sum_{j=1}^J |k_j-l_j|_0 $ with $|\cdot|_0$ being the indicator function of nonzero values.
For the BER of the MAP detector, denoted by $P_{e,b}^{\mathrm{MAP}}$, an upper bound is found to be
\begin{equation}
\label{BER_formula}
    P_{e,b} \approx P_{e,b}^{\mathrm{MAP}} \leq \frac{1}{M^J\cdot J \mathrm{log}_2M}\mathbf{1}^T_{M^J} (\mathbf{Q}\circ \mathbf{D}_b )\mathbf{1}_{M^J} 
\end{equation}
where the matrix $\mathbf{D}_b\in \mathbb{Z}^{M^J\times M^J}$ is defined with $[\mathbf{D}_b]_{k,l} = d_{b,kl}$,
 and $d_{b,kl}$ is the Hamming distance between the bit patterns loaded on the multiplexed symbols $\mathbf{m}_k, \mathbf{m}_l$, i.e., $d_{b, kl} = \sum_{j=1}^J \left\| {\bf b}^{[k_j]} - {\bf b}^{[l_j]}\right\|_0$ with $\left\|\cdot \right\|_0$ being the $\ell^0$ norm in \cite{0norm}, and ${\bf b}^{[k_j]}$, ${\bf b}^{[l_j]}\in\mathbb{B}^{\mathrm{log}_2M}$ denote the bits corresponding to symbol $k_j$, $l_j$, respectively, according to the convention in Section \ref{Sec:SCMA_Encoder}. The derivation of \eqref{BER_formula} can be done by the definitions of SER, BER, and \eqref{SER_formula}.

\begin{figure}[ht]
    \centering\centerline{
    \includegraphics[width=0.54\textwidth,clip]{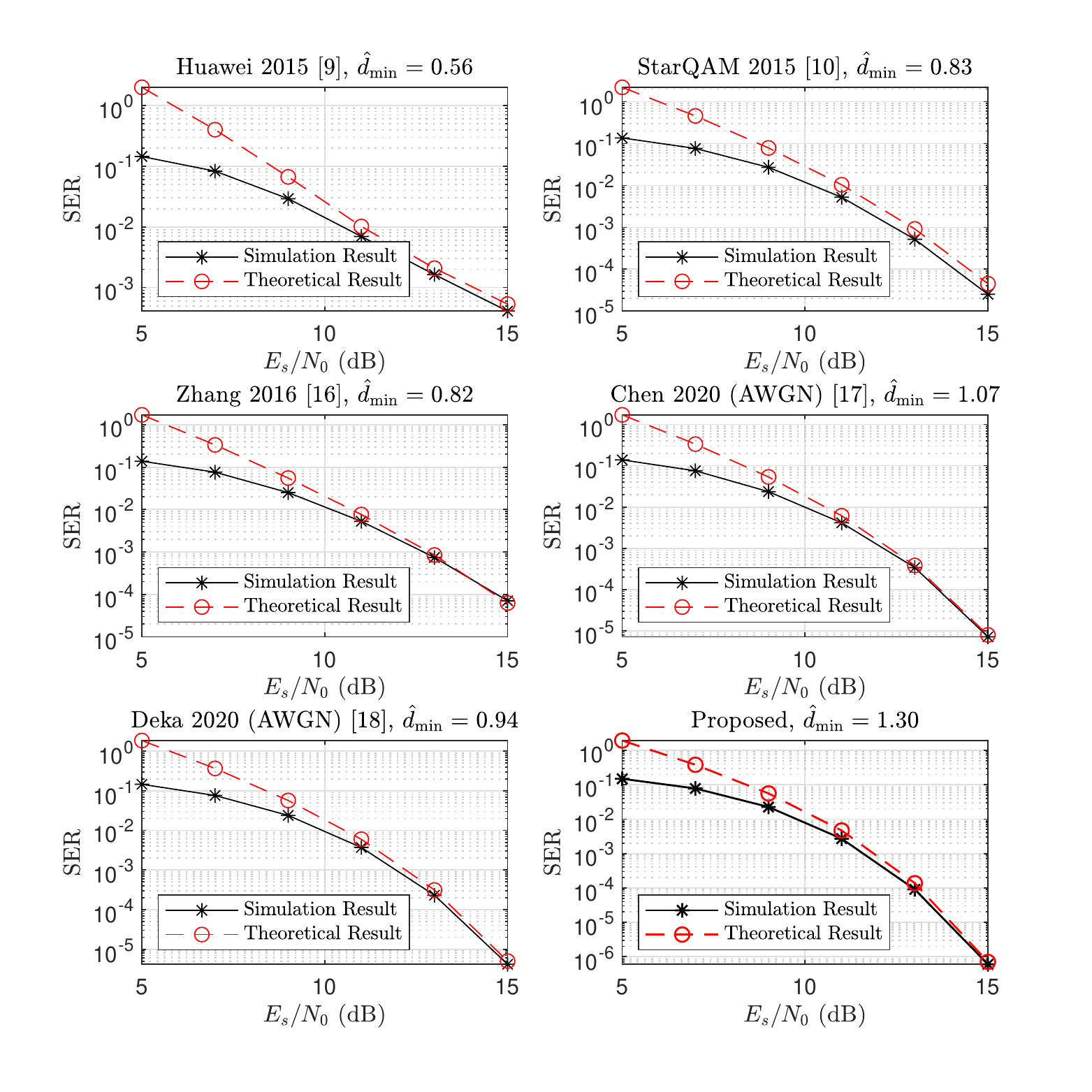}}
    \caption{Comparison of theoretical and simulation results of SER over AWGN channel.}
    \label{fig:analytic_simulation_comparison_SER}
\end{figure}

\begin{figure}[ht]
    \centering\centerline{
    \includegraphics[width=0.54\textwidth,clip]{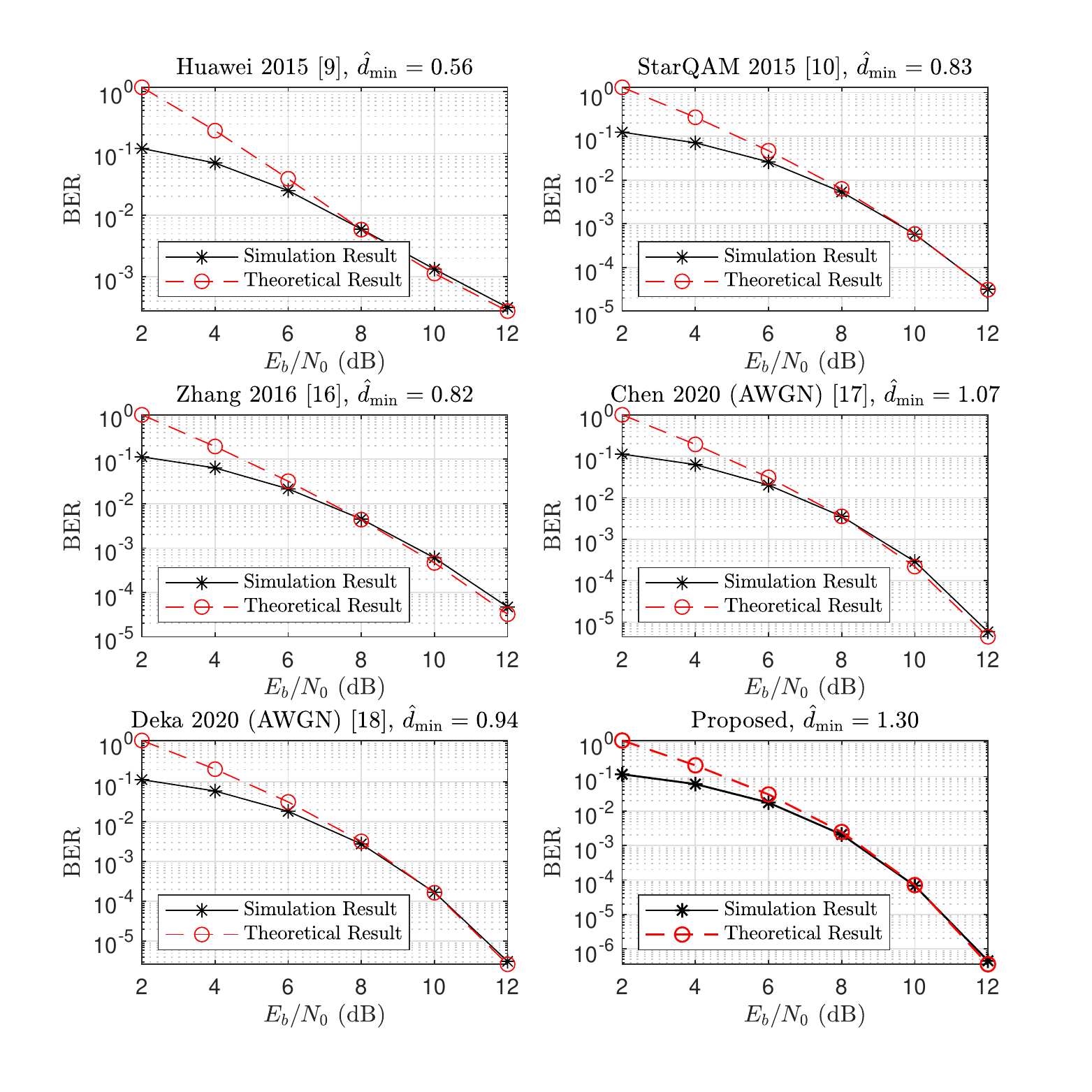}}
    \caption{Comparison of theoretical and simulation results of BER over AWGN channel.}
    \label{fig:analytic_simulation_comparison_BER}
\end{figure}

In the following comparisons shown in Fig. \ref{fig:analytic_simulation_comparison_SER}, and  \ref{fig:analytic_simulation_comparison_BER}, we will find that all simulation curves are matching the theoretical upper bounds \eqref{SER_formula}, \eqref{BER_formula} in high SNR regions within acceptable margins.
These results not only double-checked the correctness of the bounds, but also secured all performance advantages of the proposed codebook collection that we have seen in Fig. \ref{fig:SCMA_Codebook_comparision_J6_SER},  \ref{fig:SCMA_Codebook_comparision_J6_BER}, and \ref{fig:BER_simulation_fading_dicardBad}.
The bounds are rather loose in low SNR regions, but they go tighter as SNR goes higher. Sometimes it is observed that simulation results even have slightly larger error rates than the bounds, and we believe this is because the MPA is still worse than MAP.
Specifically, we notice MPA detection may not always take the closest superimposed codeword to the received signal $\mathbf{r}[b]$ as the detected one as MAP does, and this may have resulted in the slightly larger error rates than the theoretical bounds.

\begin{figure}[ht]
    \centering\centerline{
    \includegraphics[width=0.54\textwidth,clip]{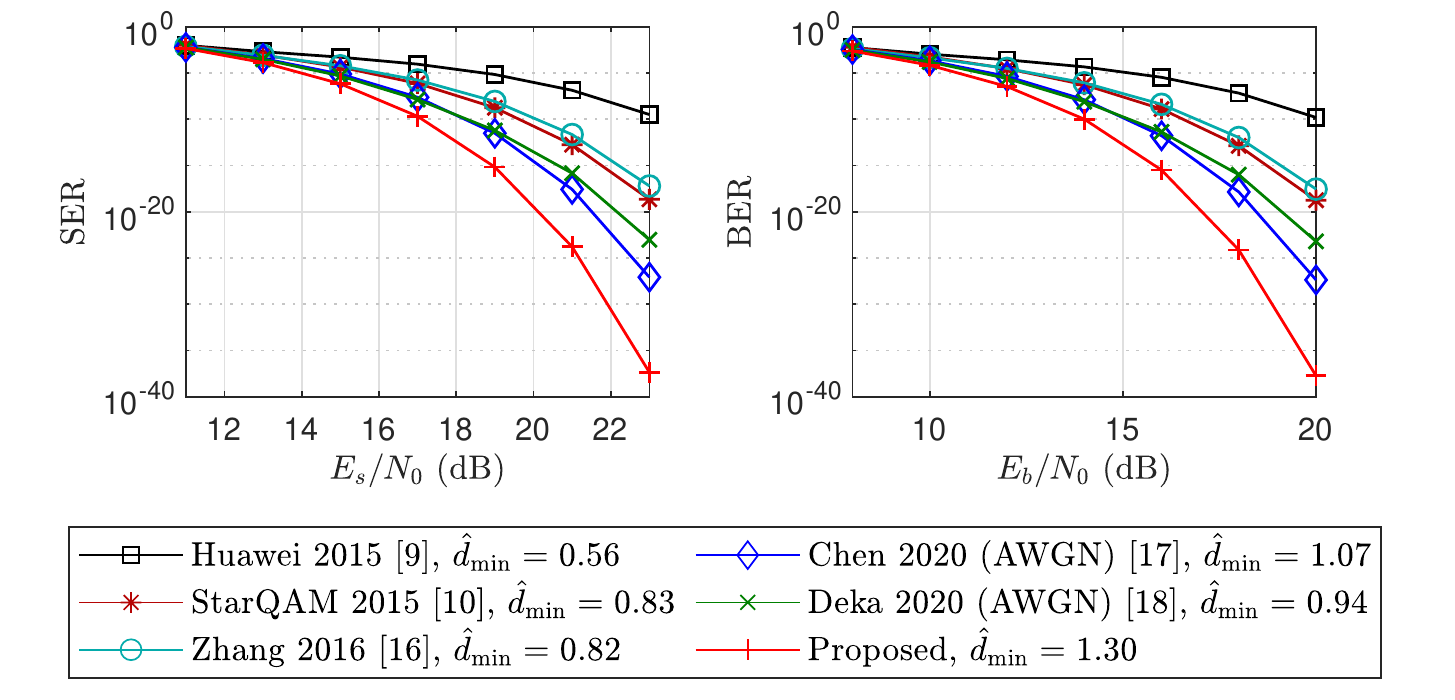}}
    \caption{Performance prediction comparisons over AWGN channel}
    \label{fig:analytic}
\end{figure}

Based on the tightness of the upper bounds and the simulation results discussed above, we can further predict the SER, and BER of some codebook collections in the high SNR region as shown in Fig. \ref{fig:analytic}. 
It can be observed that the proposed codebook collection still has the best error rate performances.

Moreover, some remarks about the relation of Euclidean distances and error rates are made. 
It is observed that the upper bounds, \eqref{SER_formula}, \eqref{BER_formula}, are both proportional to some weighted sums of all elements in $\mathbf{Q}$, so it is desirable to minimize the entries of ${\bf Q}$, which depend on the Euclidean distances, for a low error rate.
Since the largest entry in ${\bf Q}$ is $Q\left(\frac{d_{\mathrm{min}}}{\sqrt{2N_0}}\right)$, which will dominate the contributions to error rate upper bound formulas (i.e., $Q(\frac{d_{\mathrm{min}}}{\sqrt{2N_0}}) \gg Q(\sqrt{\frac{d_{kl}}{2N_0}})$ for most $k,l\in\mathcal{Z}_{M^J}, k\neq l$) when the SNR goes to infinity\footnote{This can be seen by the fact that $\displaystyle \lim_{x\rightarrow \infty} Q(ax)/Q(x) = 0$ for all $ a > 1$.} (i.e., $N_0\rightarrow 0$), our approach of maximizing $d_{\mathrm{min}}$ for the codebook collection design problem is thus justified based on the tightness of these bounds.
At least, it suggests that the codebook collection designed will have clear advantages in high-SNR regions, as we have shown in this section.







\section{Conclusions}
\label{Sec:Conclusion}

In this paper, a new method for downlink SCMA codebook design, based on maximizing the minimum Euclidean distance (MED) of superimposed codewords, is proposed. 
An iterative algorithm based on alternating maximization is applied by reformulating the MED-maximization problem into a biconvex form with an exact penalty function.
With appropriate choices of the initial codebook collection and weighting coefficients, the proposed algorithm has successfully produced a codebook collection with an MED greater than any existing codebook collection with a large margin, for the most popular six-user four-resource case.
A Lagrange dual problem of the MED-maximizing problem was derived and solved, resulting in a theoretical MED upper bound of any SCMA codebook collections that were unknown before.
Although the codebook collection reported in this article has achieved an MED that is much closer to the upper bound than any previously reported codebook collections, the fact that there is still a nonzero gap between the upper bound and the largest MED suggests there is still room for codebook improvement in the future.

Simulation results demonstrate clear advantages of the obtained largest-MED codebook collection in terms of error-rate performance over all available reported codebook collections. 
The performance advantages are not only seen in AWGN channels but also in frequency-selective fading channels assuming downlink SCMA resources are allocated from an underlying OFDMA system with resource elements with sufficiently good channel quality, which fits in the downlink scenario of 5G/NR applications, making it a promising codebook candidate in such scenarios.
The proposed codebook collection, however, does not show a performance advantage over existing ones in uplink scenarios or scenarios where non-consecutive subcarriers are used as the underlying orthogonal resources. 
The validity of all simulation curves is further verified by comparing them with theoretical error rate bounds.

In the future, it is still desirable to further reduce the duality gap between the attained MED and the upper bound obtained from the dual problem: either a codebook collection with an even larger MED is to be found, or a smaller upper bound of MED is to be derived, or both.
The choices of the initial codebook collection and weighting coefficients appear to play important roles in the proposed biconvex algorithm. 
It is therefore desirable to try out other possible combinations of these parameters to find even a better codebook collection.
\appendices
\section{Derivation of the distance matrices in \eqref{eq:Ai}}\label{Derivation_of_A_i}
First note that the difference of some pair of superimposed codewords corresponding to some two multiplexed symbols $\mathbf{m}_k,\mathbf{m}_l$ is 
{\footnotesize
\begin{align*}
\sum_{j=1}^J {\bf V}_j{\bf C}_j\left( {\bf e}^{(M)}_{k_j} - {\bf e}^{(M)}_{l_j}\right) 
&=& \sum_{j=1}^J \mathrm{vec}\left( {\bf V}_j{\bf C}_j\left( {\bf e}^{(M)}_{k_j} - {\bf e}^{(M)}_{l_j}\right)\right) \\
&=& \sum_{j=1}^J\left[ \left({\bf e}^{(M)}_{k_j} - {\bf e}^{(M)}_{l_j}\right)^T\otimes {\bf V}_j\right] {\bf x}_j
\end{align*}
}where ${\bf x}_j = \mathrm{vec}({\bf C}_j)$ is actually the $j$-th $MN$-block of ${\bf x}$. 
Therefore, 
\begin{equation}\label{x_j_in_x}
    {\bf x}_j = \left(({\bf e}_j^{(J)})^T\otimes {\bf I}_{MN}\right){\bf x}.
\end{equation}
Then the square of the Euclidean distance of this pair will be 
{\footnotesize
\begin{align*}
&\quad\enspace \left\| \sum_{j=1}^J\left[ \left({\bf e}^{(M)}_{k_j} - {\bf e}^{(M)}_{l_j}\right)^T\otimes {\bf V}_j\right] {\bf x}_j\right\|_2^2\\
&=  \sum_{q=1}^J{\bf x}_q^H \left[ \left({\bf e}^{(M)}_{k_q} - {\bf e}^{(M)}_{l_q}\right)^T\otimes {\bf V}_q\right]^H    \sum_{j=1}^J\left[ \left({\bf e}^{(M)}_{k_j} - {\bf e}^{(M)}_{l_j}\right)^T\otimes {\bf V}_j\right] {\bf x}_j\\
&= \sum_{q=1}^J\sum_{j=1}^J\mathbf{x}_q^H \left(  \left({\bf e}^{(M)}_{k_q} - {\bf e}^{(M)}_{l_q}\right)\otimes {\bf V}_q^T\right)  \left( \left({\bf e}^{(M)}_{k_j} - {\bf e}^{(M)}_{l_j}\right)^T\otimes {\bf V}_j\right)\mathbf{x}_j\\
&= \sum_{q=1}^J\sum_{j=1}^J\bigg[\mathbf{x}_q^H \left(  \left({\bf e}^{(M)}_{k_q} - {\bf e}^{(M)}_{l_q}\right) \left({\bf e}^{(M)}_{k_j} - {\bf e}^{(M)}_{l_j}\right)^T\right) \otimes \left( {\bf V}_q^T {\bf V}_j\right)\mathbf{x}_j\bigg].
\end{align*}}

In the end, by substituting \eqref{x_j_in_x} for $\mathbf{x}_j$ for all $j\in \mathcal{Z}_J$, we have the expression of $\mathbf{x}^H\mathbf{A}_i\mathbf{x}$ as shown in Section \ref{max_MED}. Note that the index $i\in \mathcal{Z}_{\binom{M^J}{2}}$ here can be mapped freely to the $\binom{M^J}{2}$ pairs of multiplexed symbols as long as the mapping is confirmed to be one-to-one.
\if01
\section{Proof of Lemma \ref{lemma:power_equality}}\label{Lemma1_proof}
For convenience, we first consider an equivalent problem of Problem \eqref{OriginProb}, which is called as Problem \eqref{QCQP_Prob}' and is defined by replacing \eqref{QCQP_Prob3} in  Problem \eqref{QCQP_Prob}  by
\begin{equation}\label{eq:equal_power_QCQP}
    \mathbf{x}^H\mathbf{B}_j\mathbf{x} \leq MP, \ \forall j \in \mathcal{Z}_J.
\end{equation}
Assume that there exists some optimal point of Problem \eqref{QCQP_Prob}',  $\{\mathbf{x}^\star, t^\star\}$, such that $\mathbf{x}^\star\mathbf{B}_{\theta}\mathbf{x}^\star=MP^\prime$ for some $\theta\in\mathcal{Z}_J$ and $P^\prime\leq P, P^\prime \neq 0$.
Then, by scaling the power of the part related to the $\theta$-th user, we can let 
\begin{equation}\label{eq:power_up}
    \mathbf{y}=\sqrt{\frac{P}{P^{\prime}}} \mathbf{v}_1+\mathbf{v}_2,
\end{equation}
where $\mathbf{v}_1=\mathbf{B}_{\theta}\mathbf{x}^\star$, $\mathbf{v}_2=\left(\mathbf{I}_{n_x}-\mathbf{B}_{\theta}\right)\mathbf{x}^\star$. 
And we have
\begin{equation*}
    \mathbf{y}^H\mathbf{B}_j\mathbf{y}=\begin{cases}\mathbf{x}^\star\mathbf{B}_j\mathbf{x}^\star, & j\neq \theta\\
MP, & j= \theta
\end{cases},
\end{equation*} which means $\mathbf{y}$ satisfies constraint \eqref{eq:equal_power_QCQP} and $\mathbf{x}^H\mathbf{B}_\theta\mathbf{x}=MP$.
Then, for any $i\in \mathcal{Z}_{\binom{M^J}{2}}$, we have
{\small
\begin{align*}
    &\mathbf{y}^H\mathbf{A}_i\mathbf{y}-\mathbf{x}^\star\mathbf{A}_i\mathbf{x}^\star \\
    &=\left(\frac{P}{P^\prime}-1\right)\mathbf{v}^H_1\mathbf{A}_i\mathbf{v}_1+2\left(\sqrt{\frac{P}{P^\prime}}-1\right)\mathrm{Re}\{\mathbf{v}^H_1\mathbf{A}_i\mathbf{v}_2\}\\
    & = \left(\sqrt{\frac{P}{P^\prime}}-1\right)\left[\left(\sqrt{\frac{P}{P^\prime}}+1\right)\mathbf{v}^H_1\mathbf{A}_i\mathbf{v}_1+2\mathrm{Re}\{\mathbf{v}^H_1\mathbf{A}_i\mathbf{v}_2\}\right]\\
    &\geq 2\left(\sqrt{\frac{P}{P^\prime}}-1\right)\left[\mathbf{v}^H_1\mathbf{A}_i\mathbf{v}_1+\mathrm{Re}\{\mathbf{v}^H_1\mathbf{A}_i\mathbf{v}_2\}\right]\\
    & = 2\left(\sqrt{\frac{P}{P^\prime}}-1\right)\left[(\mathbf{x}^\star)^H\mathbf{A}_i\mathbf{x}^\star-\mathrm{Re}\{\mathbf{v}^H_2\mathbf{A}_i\mathbf{x}^\star\}\right]\\
    & \geq 2\left(\sqrt{\frac{P}{P^\prime}}-1\right)\left[(\mathbf{x}^\star)^H\mathbf{A}_i\mathbf{x}^\star-\lvert\mathbf{v}^H_2\mathbf{A}_i\mathbf{x}^\star\rvert\right]
\end{align*}}
Furthermore, we have

some proof...

Then, we have $\mathbf{y}^H\mathbf{A}_i\mathbf{y}\geq\mathbf{x}^\star\mathbf{A}_i\mathbf{x}^\star$,
which means we can find some $t$ such that
\begin{equation*}
    t = \underset{i}{\mathrm{min}}\ \mathbf{y}^H\mathbf{A}_i\mathbf{y} \geq \underset{i}{\mathrm{min}}\ \mathbf{x}^\star\mathbf{A}_i\mathbf{x}^\star=t^\star,
\end{equation*}
and we thus have $t=t^\star$ by the assumption of optimality.

The result above means for any optimal point $\{\mathbf{x}^\star, t^\star\}$, we can always find some point $\{\mathbf{y}, t\}$ satisfying constraint \eqref{QCQP_Prob3} (and also constraint \eqref{eq:equality_power_constraint}) and $t= t^\star$ by redoing the power scaling process \eqref{eq:power_up} for all $\theta$. 
\fi
\section{The codebook collection obtained by Algorithm \ref{biconvex} with Chen's AWGN codebook collection\cite{Chen2020} for initialization}\label{Chen_init}
$\mathcal{S}_1-\mathcal{S}_6$, shown as follows.
\begin{center}
\resizebox{\columnwidth}{!}{
\begin{tabular}{ l } 
 $\mathcal{S}_1= 
     \begin{Bmatrix}
     \begin{bmatrix} 
     -0.9643 - 0.0000i \\ -0.0698 - 0.2553i \\ 0 \\ 0
     \end{bmatrix}
     \begin{bmatrix} 
     -0.2562 - 0.0664i \\ 0.4795 + 0.8367i \\ 0 \\ 0
     \end{bmatrix}
     \begin{bmatrix} 
     0.2562 + 0.0664i \\-0.4795 - 0.8367i \\ 0 \\ 0
     \end{bmatrix}
     \begin{bmatrix} 
     0.9643 - 0.0000i \\ 0.0698 + 0.2553i \\ 0 \\ 0
     \end{bmatrix}
     \end{Bmatrix}$ \\ \\
 $\mathcal{S}_2= 
     \begin{Bmatrix}
     \begin{bmatrix} 
     0 \\ 0 \\ -0.3519 + 0.8978i \\ -0.2121 - 0.1584i
     \end{bmatrix}
     \begin{bmatrix} 
     0 \\ 0 \\ -0.0317 + 0.2628i \\ 0.6031 + 0.7525i
     \end{bmatrix}
     \begin{bmatrix} 
     0 \\ 0 \\ 0.0317 - 0.2628i \\  -0.6031 - 0.7525i
     \end{bmatrix}
     \begin{bmatrix} 
     0 \\ 0 \\ 0.3519 - 0.8978i \\ 0.2121 + 0.1584i
     \end{bmatrix}
     \end{Bmatrix}$ \\ \\
 $\mathcal{S}_3= 
     \begin{Bmatrix}
     \begin{bmatrix} 
     -0.2821 + 0.8866i \\ 0 \\ -0.3653 + 0.0320i \\ 0
     \end{bmatrix}
     \begin{bmatrix} 
     -0.1631 + 0.3284i \\ 0 \\ 0.9284 + 0.0609i \\ 0
     \end{bmatrix}
     \begin{bmatrix} 
     0.1631 - 0.3284i \\ 0 \\ -0.9284 - 0.0609i \\ 0
     \end{bmatrix}
     \begin{bmatrix} 
     0.2821 - 0.8866i \\ 0 \\ 0.3653 - 0.0320i \\ 0
     \end{bmatrix}
     \end{Bmatrix}$ \\ \\
 $\mathcal{S}_4= 
     \begin{Bmatrix}
     \begin{bmatrix} 
     0 \\ -0.9095 + 0.1960i \\ 0 \\ -0.1543 + 0.3326i
     \end{bmatrix}
     \begin{bmatrix} 
     0 \\ -0.3660 + 0.0218i \\ 0 \\ 0.5154 - 0.7746i
     \end{bmatrix}
     \begin{bmatrix} 
     0 \\ 0.3660 - 0.0218i \\ 0 \\ -0.5154 + 0.7746i
     \end{bmatrix}
     \begin{bmatrix} 
     0 \\ 0.9095 - 0.1960i \\ 0 \\ 0.1543 - 0.3326i
     \end{bmatrix}
     \end{Bmatrix}$ \\ \\
 $\mathcal{S}_5= 
     \begin{Bmatrix}
     \begin{bmatrix} 
     -0.4802 - 0.8223i \\ 0 \\ 0 \\ -0.2843 - 0.1112i
     \end{bmatrix}
     \begin{bmatrix} 
     -0.2646 - 0.1522i \\ 0 \\ 0 \\ 0.9420 - 0.1395i
     \end{bmatrix}
     \begin{bmatrix} 
     0.2646 + 0.1522i \\ 0 \\ 0 \\ -0.9420 + 0.1395i
     \end{bmatrix}
     \begin{bmatrix} 
     0.4802 + 0.8223i \\ 0 \\ 0 \\ 0.2843 + 0.1112i
     \end{bmatrix}
     \end{Bmatrix}$ \\ \\
 $\mathcal{S}_6= 
     \begin{Bmatrix}
     \begin{bmatrix} 
     0 \\ -0.4747 + 0.8255i \\ -0.0452 - 0.3019i \\ 0
     \end{bmatrix}
     \begin{bmatrix} 
     0 \\  -0.0006 + 0.3052i \\ 0.5903 + 0.7472i \\ 0
     \end{bmatrix}
     \begin{bmatrix} 
     0 \\ 0.0006 - 0.3052i \\ -0.5903 - 0.7472i \\ 0
     \end{bmatrix}
     \begin{bmatrix} 
     0 \\ 0.4747 - 0.8255i \\ 0.0452 + 0.3019i \\ 0
     \end{bmatrix}
     \end{Bmatrix}$ 
\end{tabular}

}
\end{center}
\section{The proposed codebook collection}\label{Deka_init}
$\mathcal{S}_1-\mathcal{S}_6$, shown as follows.
\begin{center} 
\resizebox{\columnwidth}{!}{
\begin{tabular}{ l } 
 $\mathcal{S}_1= 
     \begin{Bmatrix}
     \begin{bmatrix} 
     -0.4969 - 0.0000i \\ 0.2516 + 0.8044i \\ 0 \\ 0
     \end{bmatrix}
     \begin{bmatrix} 
     -0.5790 - 0.0043i \\ -0.7819 - 0.3102i \\ 0 \\ 0
     \end{bmatrix}
     \begin{bmatrix} 
     0.5790 + 0.0043i \\ 0.7819 + 0.3102i \\ 0 \\ 0
     \end{bmatrix}
     \begin{bmatrix} 
     0.4969 + 0.0000i \\ -0.2516 - 0.8044i \\ 0 \\ 0
     \end{bmatrix}
     \end{Bmatrix}$ \\ \\
 $\mathcal{S}_2= 
     \begin{Bmatrix}
     \begin{bmatrix} 
     0 \\ 0 \\ 0.3036 - 0.2955i \\ 0.6048 + 0.7008i
     \end{bmatrix}
     \begin{bmatrix} 
     0 \\ 0 \\ 0.7086 - 0.5300i \\ -0.3028 - 0.2982i
     \end{bmatrix}
     \begin{bmatrix} 
     0 \\ 0 \\ -0.7086 + 0.5300i \\  0.3028 + 0.2982i
     \end{bmatrix}
     \begin{bmatrix} 
     0 \\ 0 \\ -0.3036 + 0.2955i \\ -0.6048 - 0.7008i
     \end{bmatrix}
     \end{Bmatrix}$ \\ \\
 $\mathcal{S}_3= 
     \begin{Bmatrix}
     \begin{bmatrix} 
     0.1787 - 0.7137i \\ 0 \\ -0.5264 - 0.0696i \\ 0
     \end{bmatrix}
     \begin{bmatrix} 
     0.8996 - 0.3570i \\ 0 \\ 0.4898 - 0.0103i \\ 0
     \end{bmatrix}
     \begin{bmatrix} 
     -0.8996 + 0.3570i \\ 0 \\ -0.4898 + 0.0103i \\ 0
     \end{bmatrix}
     \begin{bmatrix} 
     -0.1787 + 0.7137i \\ 0 \\ 0.5264 + 0.0696i \\ 0
     \end{bmatrix}
     \end{Bmatrix}$ \\ \\
 $\mathcal{S}_4= 
     \begin{Bmatrix}
     \begin{bmatrix} 
     0 \\ -0.5491 + 0.0077i \\ 0 \\ 0.3049 - 0.8077i
     \end{bmatrix}
     \begin{bmatrix} 
     0 \\ 0.5034 + 0.0712i \\ 0 \\ 0.7996 - 0.2351i
     \end{bmatrix}
     \begin{bmatrix} 
     0 \\ -0.5034 - 0.0712i \\ 0 \\ -0.7996 + 0.2351i
     \end{bmatrix}
     \begin{bmatrix} 
     0 \\ 0.5491 - 0.0077i \\ 0 \\ -0.3049 + 0.8077i
     \end{bmatrix}
     \end{Bmatrix}$ \\ \\
 $\mathcal{S}_5= 
     \begin{Bmatrix}
     \begin{bmatrix} 
     0.2927 + 0.8221i \\ 0 \\ 0 \\ -0.5292 - 0.0245i
     \end{bmatrix}
     \begin{bmatrix} 
     -0.7798 - 0.2581i \\ 0 \\ 0 \\ -0.5316 - 0.0226i
     \end{bmatrix}
     \begin{bmatrix} 
     0.7798 + 0.2581i \\ 0 \\ 0 \\ 0.5316 + 0.0226i 
     \end{bmatrix}
     \begin{bmatrix} 
     -0.2927 - 0.8221i \\ 0 \\ 0 \\ 0.5292 + 0.0245i
     \end{bmatrix}
     \end{Bmatrix}$ \\ \\
 $\mathcal{S}_6= 
     \begin{Bmatrix}
     \begin{bmatrix} 
     0 \\ 0.4950 - 0.4726i \\ 0.4539 + 0.5691i \\ 0
     \end{bmatrix}
     \begin{bmatrix} 
     0 \\  0.5690 - 0.5140i \\ -0.4126 - 0.4934i \\ 0
     \end{bmatrix}
     \begin{bmatrix} 
     0 \\ -0.5690 + 0.5140i \\ 0.4126 + 0.4934i \\ 0
     \end{bmatrix}
     \begin{bmatrix} 
     0 \\ -0.4950 + 0.4726i \\ -0.4539 - 0.5691i \\ 0
     \end{bmatrix}
     \end{Bmatrix}$ 
\end{tabular}

}
\end{center}

\section*{Acknowledgment}
\noindent This work was supported by the Ministry of Science and Technology of Taiwan under Grant MOST 109-2221-E-002-155.



\end{document}